\documentclass{nature}
\usepackage{graphicx}
\usepackage{pdfpages}
\usepackage[section]{placeins}
\usepackage{amssymb,amsmath}

\usepackage{comment}
\usepackage{float}
\usepackage{caption}
\usepackage{capt-of}
\usepackage{epstopdf}
\usepackage{xcolor}
\usepackage[normalem]{ulem}
\usepackage{multirow,rotating} 
\usepackage{color}

\RequirePackage[normalem]{ulem} 
\RequirePackage{color}\definecolor{RED}{rgb}{1,0,0}\definecolor{BLUE}{rgb}{0,0,1} 

\title{Emergence of Scaling in Complex Substitutive Systems}

\author{Ching Jin$^{1,2,3*}$, Chaoming Song$^{4*}$, Johannes Bjelland$^5$, Geoffrey Canright$^5$ \& Dashun Wang$^{1,2\dagger}$}
\begin{document}

\maketitle

\begin{affiliations}
\item Northwestern Institute on Complex Systems (NICO), 600 Foster Street, Evanston, IL 60208, USA
\item Kellogg School of Management, 2211 Campus Drive, Evanston, IL 60208, USA
\item Center for Complex Network Research, Northeastern University, Boston, MA 02115, USA
\item Department of Physics, University of Miami, 1320 S Dixie Hwy, Coral Gables, FL 33146, USA
\item Telenor Research and Development, Snar{\o}yveien 30 N-1360 Fornebu, Norway

\item[*] These authors contributed equally to this work.
\item[$^\dagger$] Correspondence should be addressed to D.W. (dashun.wang@kellogg.northwestern.edu)
\end{affiliations}
\newpage

\textbf{\textit{Abstract}}. 
Diffusion processes are central to human interactions. One common prediction of the current modeling frameworks is that initial spreading dynamics follow exponential growth. Here, we find that, ranging from mobile handsets to automobiles, from smart-phone apps to scientific fields, early growth patterns follow a power law with non-integer exponents. We test the hypothesis that mechanisms specific to substitution dynamics may play a role, by analyzing a unique data tracing 3.6M individuals substituting for different mobile handsets. We uncover three generic ingredients governing substitutions, allowing us to develop a minimal substitution model, which not only explains the power-law growth, but also collapses diverse growth trajectories of individual constituents into a single curve. These results offer a mechanistic understanding of power-law early growth patterns emerging from various domains and demonstrate that substitution dynamics are governed by robust self-organizing principles that go beyond the particulars of individual systems. 
\newpage

Diffusion processes impact broad aspects of human society\cite{barrat2008dynamical, ben2000diffusion, pastor2015epidemic, rogers2010diffusion, gladwell2006tipping}, ranging from the spread of biological viruses\cite{pastor2015epidemic, anderson1992infectious, colizza2006role, brockmann2013hidden} to the adoption of innovations\cite{rogers2010diffusion, bass1969new, fisher1972simple, banerjee2013diffusion, karsai2014complex, aral2009distinguishing, weiss2014adoption} and knowledge\cite{merton1973sociology, evans2011metaknowledge} and  
to the spread of information\cite{granovetter1973strength, onnela2007structure,pentland2015social}, cultural norms and social behavior\cite{christakis2007spread, centola2010spread, morone2015influence, castellano2009statistical}. 
Despite numerous studies that span multiple disciplines, 
our knowledge is mainly limited to spreading processes in non-substitutive systems. 
Yet, a considerable number of ideas, products and behaviors spread by substitution---to adopt a new one, agents often need to give up an existing one. 
For example, the development of science 
hinges on scientists' relentlessness in abandoning a scientific  
framework once one that offers a better description of reality emerges\cite{kuhn1996structure}.
The same is true for adopting a new healthy habit or other durable items, like mobile phones, cars or homes.

While substitutions play a key role from science to economy, 
our limited understanding of such processes stems from the lack of empirical data tracing their characteristics.
To study the dynamics of substitutions, we explore growth patterns 
in four different substitutive systems where detailed dynamical patterns are captured with fine temporal resolution
(See Supplementary Note~1 for detailed data descriptions).
Our first dataset captures, with daily resolution, 3.6 Million individuals
choosing among different types of mobile handsets, 
recorded by a Northern European telecommunication company from January 2006 to November 2014. 
Since an individual is unlikely to keep more than one mobile phone at a time, 
his or her adoption of a new handset is typically associated with discontinuance of the old one. 
Here, we focus on handsets that have been released for at least 6 months and used by at least 50 users in total (885 different handset models). 
Our second dataset captures monthly transaction records of 126 automobiles sold in the North America between 2010 and 2016.   
These automobiles have been released for at least four months before the data was collected. 
Automobiles represent a similar example as mobile handsets, where adoptions are largely driven by substitutions, 
given the limited number of automobiles a typical household may have.

While handset and automobile adoptions are relatively exclusive, 
in reality, there are also ``hybrid'' substitutive systems, where the definition of substitutions is less strict. 
To test if results presented in this paper may apply to such systems, we collected two additional datasets: 
One traces the number of daily downloads for new smartphone apps published in the App store 
(2,672 most popular apps in the iOS systems from November to December 2016), 
and the other one is a scientific publication dataset, recording 246,630 scientists substituting for 6,399 scientific fields from 1980 to 2018.
Indeed, usages of smart-phone apps are subject to constraints of time and device space, 
hence a new app downloaded reduces the usage of other similar apps, if not replacing them all together.
Yet, at the same time, apps may also be downloaded without involving substitutions. 
Similarly, while many scientists may focus on one research area at a time\cite{jia2017quantifying}, 
where research direction shifts may be characterized by substitutions, there 
are also people who explore several directions simultaneously hence an increased focus on one direction does not necessarily imply a decreased attention to others.

\textbf{\textit{Results}}.
 
\begin{figure} 
\centering
\resizebox{1.0\hsize}{!}
{\includegraphics{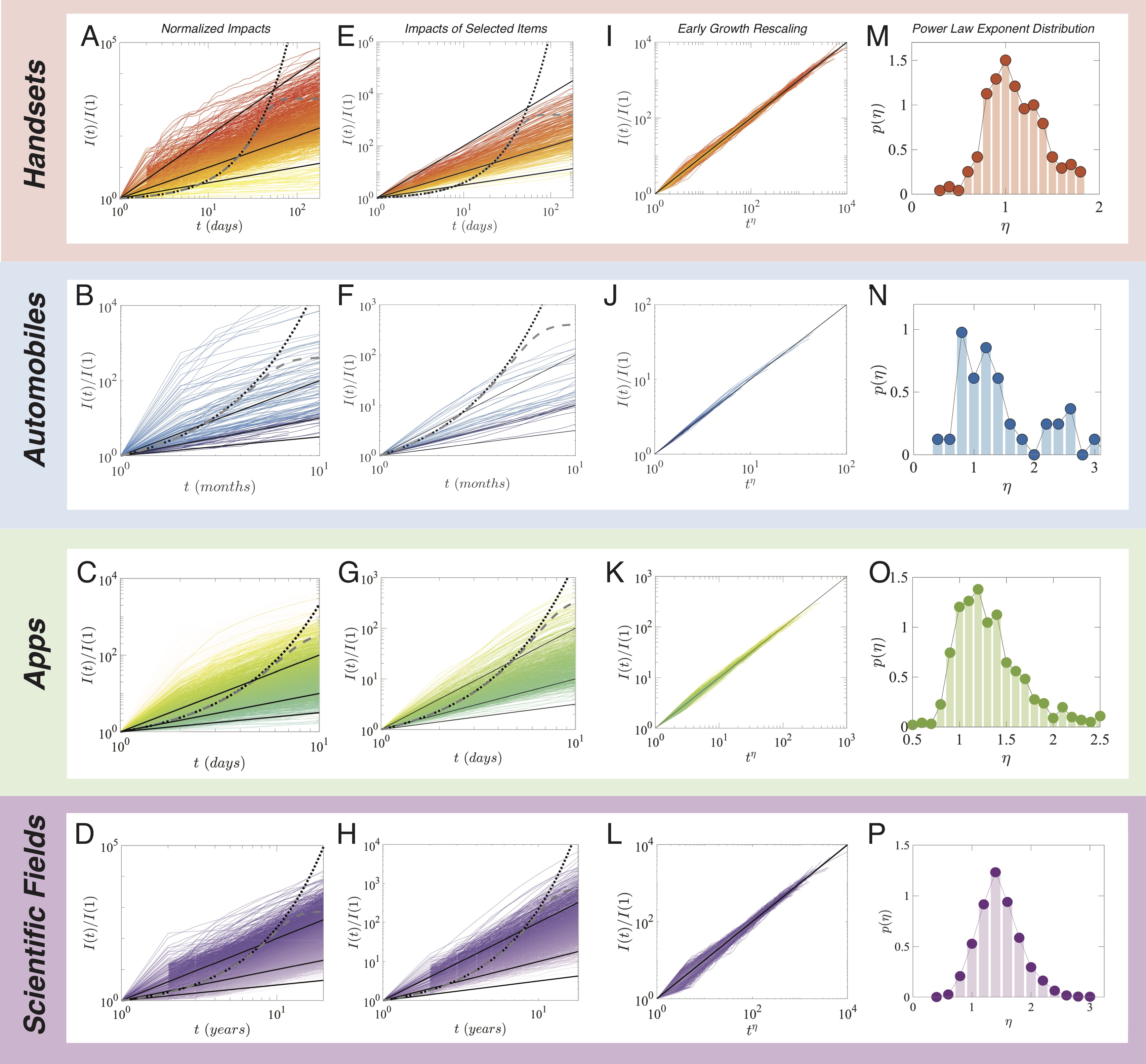}}
\caption{
} 
\label{fig:early}
\end{figure}

A common prediction by current modeling frameworks, from epidemiological models\cite{pastor2015epidemic,anderson1992infectious} to disordered systems\cite{ben2000diffusion} 
to diffusion of innovations\cite{rogers2010diffusion,karsai2014complex}, 
is that early growth patterns follow an exponential function. 
To test this prediction, we measure in our four datasets the impact of each mobile handset, automobile model, smartphone app, and scientific field. 
More specifically, we calculated  $I(t)$, 
measuring the number of individuals who bought the handset up to time $t$ since its availability (Fig.~\ref{fig:early}A), 
cumulative sales of an automobile (Fig.~\ref{fig:early}B), 
daily downloads of an App (Fig.~\ref{fig:early}C), 
and the number of publishing scientists in a field (Fig.~\ref{fig:early}D), respectively. 
To compare across different constituents, we normalized $I(t)$ by its initial value $I(1)$ (i.e. the first day or year when the constituent was introduced), 
and first focused on their early growth periods only (Supplementary Note~1).

We find that, in contrast to the exponential curves predicted by canonical models, 
for many of the constituents across the four systems, their growth trajectories appear to follow straight lines on a log-log plot (Fig.~1A--D),
suggesting that they may be described by power law functions.
This observation prompts us to systematically test whether power law or exponential-class functions (exponential or logistic) are preferred to describe early growth curves observed in our four systems.  
Using the Akaike information criterion (AIC), we find that 98.6\% handsets, 83.5\% automobiles, 79.6\% apps and 74.1\% scientific fields favor power-law early growth patterns (Supplementary Note~1 and Supplementary Figure~12). 
We further tested the robustness of this result by applying different statistical tests (Supplementary Note~1), and by varying the definition of early growth periods in each dataset (Supplementary Figure~14), and for both cases, we arrived at the same conclusion. 

Note that, although for vast majority of the curves (80.18\%--99.21\%), power law offers a better fit than exponential-class models (see Supplementary Note~1, Supplementary Figure 4, 12 and 14), 
there is variability in how well a power law function fits different curves. 
Moreover, there is a small fraction (0.79\%--19.82\%) of constituents whose early growth patterns can be described by exponential functions, suggesting that for these constituents their growth patterns are consistent with the predictions of existing models. 
To ensure our fitting procedure is not biased against exponential functions, we analyzed spreading patterns of 168 cases of flu pandemics in the United States, where early growth patterns are expected to follow exponential function. We find the fitting results indeed systematically prefer exponential to power law (Supplementary Note~1 and Supplementary Figure~15). 
Together, Fig.~1A-D suggest the existence of a non-trivial fraction (74.1\%--98.6\%) of constituents, whose early growth patterns follow a power law rather than an exponential function.

To examine if there are indeed a fraction of growth trajectories that can be well described by power law growth patterns, 
we further restrict the criteria for classifying power laws by selecting on those with a high $R^2$ in fitting (e.g. $R^2>0.99$). 
We find that, under the stricter criteria, a substantial fraction of constituents remained in each of the four systems (27.12\% handsets, 29.37\% automobiles, 38.25\% Apps, and 27.24\% fields) (Fig.~1E--H).
The results indicate that for a substantial fraction of constituents across the four substitutive systems we studied, their impacts grow following
\begin{equation}
I(t)/I(1) = t^{\eta_i}.  
\label{eq:Itau}
\end{equation}

We also noticed that within each system, the slopes of power-law curves shown in Fig.~1E--H differ across different constituents, suggesting that each of them is characterized by constituent-specific exponents ($\eta_i$). 
To test this hypothesis, we plot each curve in Fig.~\ref{fig:early}E--H in terms of  $t^{\eta_i}$.
As constituents differ from each other, the rescaled curves show variations around the function $y=x$. 
Yet we find that most curves are reasonably collapsed onto the same function (Fig.~\ref{fig:early}I--L). 
The rescaled growth patterns for all products across our four datasets are also shown in Supplementary Figure~6. We find that, although as expected, their growth patterns show more variations around $y=x$, they are clearly different from exponential growth patterns.

The observations documented in Fig.~1A--L are somewhat unexpected for two main reasons. 
First, the four systems we studied differ widely in their scope, scale, temporal resolution, and user demographics. 
Yet, we find, independent of the nature of the system and the identity of the constituents, 
their early growth follows similar patterns, showing that a power law scaling emerges across all four systems. 
Second, exponents $\eta_i$ are mostly non-integers (Fig.~\ref{fig:early}M--P). 
Power law growth with such non-integer exponents is rare because it corresponds to non-analytic behavior. Indeed,
due to the inability to express them in terms of taylor series around $t=0$, power laws with non-integer exponents indicate singular behavior around the release time (the $\lceil{\eta}\rceil$-th order derivative diverges at $t=0$). 
Current modeling frameworks\cite{pastor2015epidemic,anderson1992infectious, ben2000diffusion, rogers2010diffusion,karsai2014complex} 
rely on functions without singularities, hence are unable to anticipate non-analytic solutions 
(Detailed descriptions and comparisons to existing models are described in Supplementary Note~1 and 3). 
Indeed, comparing with exponential growth, power law encodes an early divergence, corresponding to an explosive growth 
at the moment when new constituents are introduced. 
Yet following this brief singularity,
the number of users grows much more slowly than what exponential functions predict, suggesting that substitutive innovations spread more slowly beyond the initial excitement.

Keep in mind, however, that not all curves follow power law growth patterns, and a few of them can indeed be described by exponential functions, suggesting that substitutions and traditional adoptions may coexist in our systems. 
Nevertheless, these results document the existence of power law early growth curves in the substitutive systems we studied, 
a pattern that is not anticipated by traditional modeling frameworks, and suggests that substitutive systems may be governed by different dynamics.

To be sure, power laws can be generated in real networks due to the growth of the systems\cite{barabasi2016network, zang2016beyond}. 
To check if Fig.~{\ref{fig:early}} may be explained by gradual addition of new users to the underlying network,  
we removed new mobile subscribers in the mobile-phone dataset and measured again $I(t)$ for different handsets.
We find that the power law scaling holds the same (Supplementary Note~3, Supplementary Figure~19), 
indicating that the scaling observed in Fig.~{\ref{fig:early}} is governed by mechanisms that operate \emph{within} the system, not driven by growth of the system. 
Another possible origin of power law growth is rooted in the bursty nature of human behavior\cite{song2010limits,kooti2016portrait}, 
where the inter-event time between adoptions follows a power law distribution. 
We measured this quantity directly in the mobile-phone dataset, finding the data systematically reject power law as a viable function to describe the inter-event time distribution ($p<10^{-3}$, Supplementary Figure~20).
It is also worth noting that, sub-exponential growth patterns have recently been found in the spread of epidemics such as Ebola and HIV\cite{chowell2015western, chowell2016mathematical, chowell2017perspectives}. 
There are also phenomenological models of spreading dynamics that take power law early growth as their assumptions\cite{chowell2016mathematical,danon2016need, chowell2016early}, in addition to a large body of literature
on modeling popularity dynamics\cite{wu2007novelty, crane2008robust, iribarren2009impact, gleeson2016effects, gleeson2014simple, shen2014modeling, wang2013quantifying}. While a mechanistic explanation is still lacking,
these examples demonstrate that the power-law early growth patterns uncovered here may hold relevance to a broad array of areas. 
Together, these results raise a fundamental question: what is the origin of the power law growth pattern?

\begin{figure} 
\centering
\resizebox{1.1\hsize}{!}
{\includegraphics{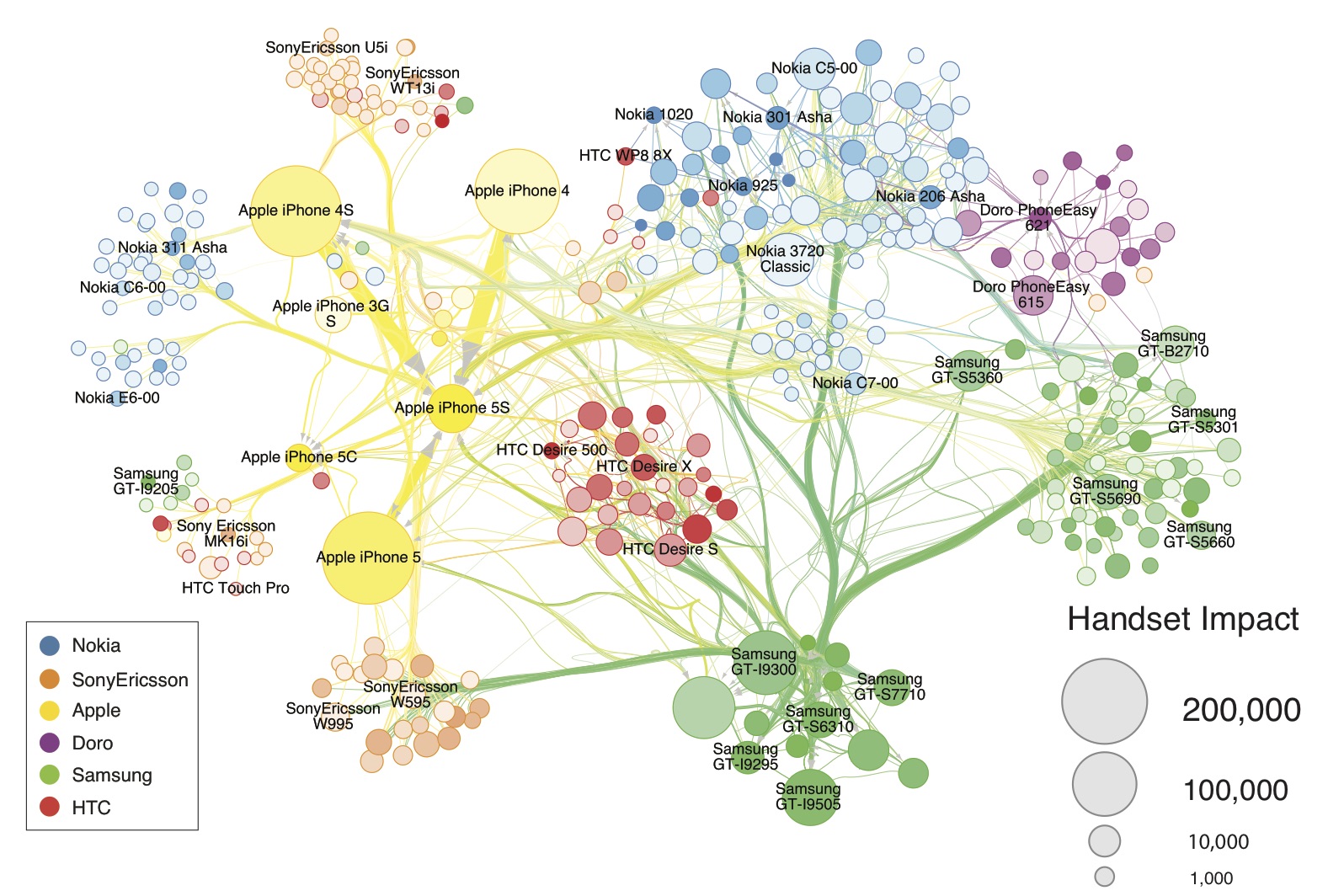}}
\caption{
}

\label{fig:network}
\end{figure}

\begin{figure} 
\centering
\resizebox{0.98\hsize}{!}
{\includegraphics{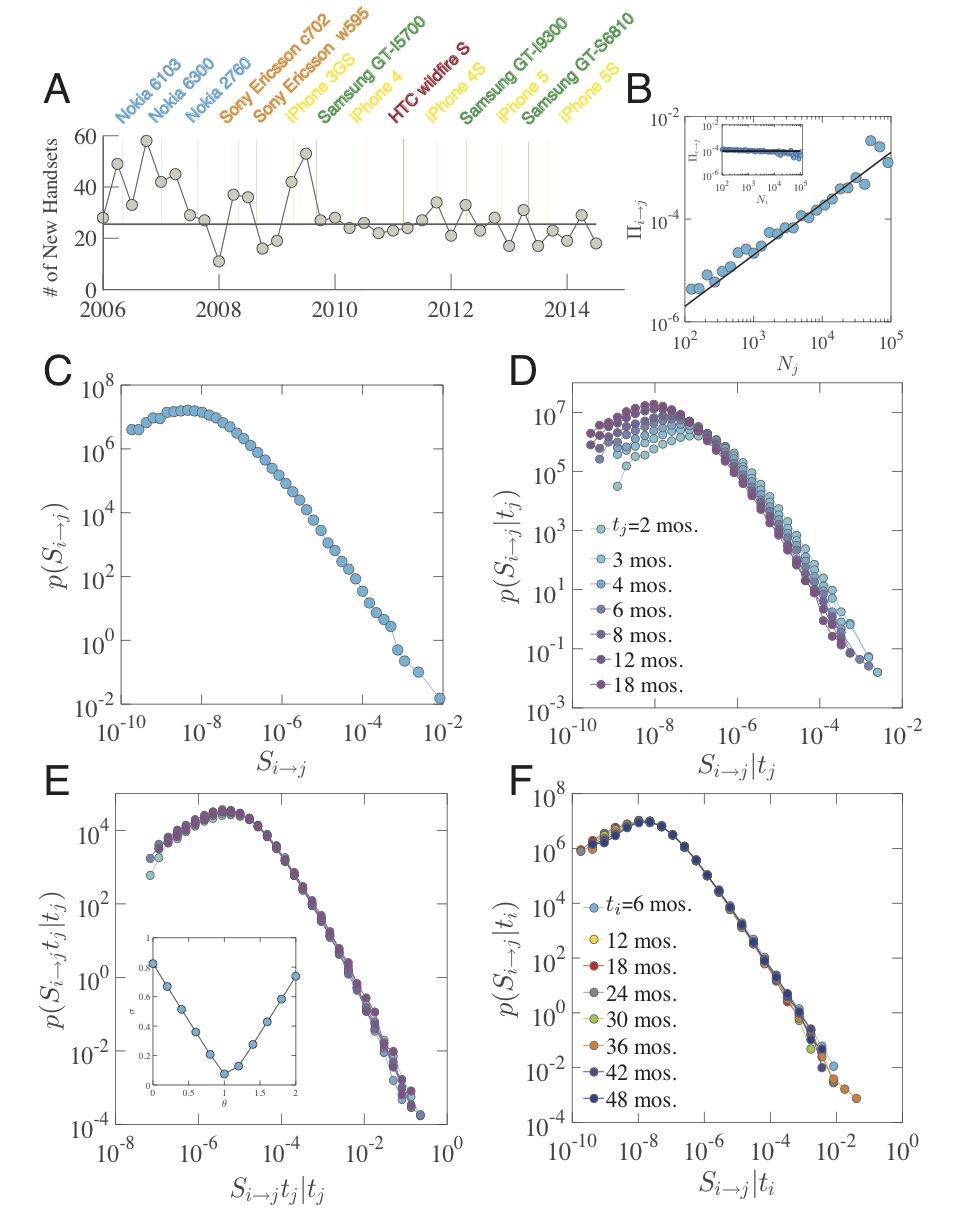}}
\caption{
} 
\label{fig:basics}
\end{figure}

\textbf{\textit{Quantifying substitution patterns}}. 
A common characteristic of the four studied systems is that they evolve by substitutions.  
In this respect, mobile phones represent an ideal setting 
for the empirical investigation of substitutive processes. 
Indeed, each time a user purchases a handset, the transaction history is recorded by telecommunication companies. 
Anonymized phone numbers together with their portability across devices provide individual traces for substitutions. 
We examined detailed user histories in the mobile-phone dataset, finding that the adoption and discontinuance histories are indeed predominantly represented by substitutions (Supplementary Note~2). 
Each type of handset is substituted by a large number of other handsets, 
hence substitution patterns are characterized by a dense, heterogeneous network that evolves rapidly over time ($\left<k\right>\approx73.6$, Supplementary Figure~17BC and 18).
To visualize substitution patterns, we applied a backbone extraction method\cite{serrano2009extracting} to identify statistically significant substitution flows for each handset given its total substitution volumes (Fig.~\ref{fig:network}). 
While mobile handsets have changed substantially over the years, 
undergoing a ubiquitous shift from feature phones to smart phones, 
the rate at which new handsets enter the market remained remarkably stable (Fig.~\ref{fig:basics}A), 
highlighting the highly competitive nature of the system:
Ensuing generations of new handsets enter the market in a somewhat regular manner, substituting for the incumbent, 
thereby affecting the rise and fall of their popularities
(Supplementary Figure~18A).

To uncover the mechanisms governing substitution dynamics,
we note that the rate of change in $N_{i}(t)$, the number of users for handset $i$ at time $t$, can be expressed in terms of 
the probability for individuals to transition from all other handsets ($k$) to $i$, $\Pi_{k \rightarrow i}$, subtracted by those leaving $i$ for other handsets ($j$), $\Pi_{i \rightarrow j}$:
\begin{equation}
\frac{d N_{i}(t)} { d t}  =  \sum_{k} \Pi_{k \rightarrow i}(t) N_{k}(t)- \sum_{j} \Pi_{i \rightarrow j}(t) N_i(t).
\label{eq:masterEq}
\end{equation}
The key to solving the master equation (\ref{eq:masterEq}) is to determine $\Pi_{i \rightarrow j}$, the substitution probability for a user to substitute handset $i$ for $j$ at time $t$.
As we show next, $\Pi_{i \rightarrow j}$ is driven by three mechanisms: \emph{preferential attachment}, \emph{recency} and \emph{propensity}. 

Figure~\ref{fig:basics}B shows that $\Pi_{i \rightarrow j}$ is independent of the number of individuals using $i$ ($N_{i}$), but proportional to $N_{j}$: $\Pi_{i \rightarrow j}\sim N_{j}$. 
This result captures the well-known preferential attachment effect\cite{merton1973sociology, barabasi2016network}: More popular handsets are more likely to attract new users than their less popular counterparts, consistent with existing models that can be used to characterize substitutions\cite{lotka1910contribution, volterra1928variations}. 
Yet $N_{j}$ by itself is insufficient to explain $\Pi_{i \rightarrow j}$. 
Indeed, we further normalized $\Pi_{i \rightarrow j}$ by $N_{j}$, by defining $S_{i \rightarrow j} \equiv \Pi_{i \rightarrow j}/N_{j}$, 
the substitution rate at which handset $j$ substitutes for $i$. 
We find that $p(S_{i \rightarrow j})$ follows a fat-tailed distribution spanning several orders of magnitude (Fig.~\ref{fig:basics}C), indicating that 
substitution rates are characterized by a high degree of heterogeneity, 
where $S_{i \rightarrow j}$ between some handset pairs are orders of magnitude higher than others.

To identify mechanisms responsible for the observed heterogeneity in $S_{i \rightarrow j}$, we grouped  $S_{i \rightarrow j}$ based on the age of the substitutes $t_j$, the number of days elapsed since its release date,  
and measure the conditional probability $p(S_{i \rightarrow j}|t_j)$ for each group. 
We find that as substitutes grow older (increasing $t_j$), 
$p(S_{i \rightarrow j}|t_j)$ shifts systematically to the left (Fig.~\ref{fig:basics}D), 
indicating substitution rates decrease with the age of substitutes---newer handsets substitute for the incumbents at a higher rate. 
Yet, within each group, the heterogeneity of $S_{i \rightarrow j}$ persisted, as $p(S_{i \rightarrow j}|t_j)$ again follows a fat-tailed distribution.
Once we rescale the distributions $p(S_{i \rightarrow j}|t_j)$ with $t_j$, however, we find that
all seven distributions  
in Fig.~\ref{fig:basics}D collapse into one single curve (Fig.~\ref{fig:basics}E).
To quantify the relationship between $S_{i \rightarrow j}$ and $t_j$, we take an ansatz: 
$S_{i \rightarrow j} \sim t_j^{-\theta}$, and rescale $S_{i \rightarrow j}$ by $t_j^{-\theta}$. 
As we vary $\theta$, we monitor the diversity of the curves, finding that it reaches its minimum around $\theta=1$ (Fig.~\ref{fig:basics}E, inset), 
indicating  $S_{i \rightarrow j}$ is inversely proportional to $t_j$. 
The data collapse in Fig.~\ref{fig:basics}E demonstrates that a single distribution characterizes substitution rates, independent of the age of substitutes:
\begin{equation}
p(S_{i \rightarrow j}|t_j)\sim t_j\mathcal{F}(S_{i \rightarrow j}t_j).
\label{eq:collapse_tau_j}
\end{equation}
In other words, substitution rates $S_{i \rightarrow j}$ can be decomposed into two independent factors: one is the universal function $\mathcal{F}(x)$, which is independent of the substitute's age, capturing an inherent propensity-based heterogeneity among handsets. 
Denoting the propensity by $\lambda_{ij}\equiv S_{i \rightarrow j}t_j$, (\ref{eq:collapse_tau_j}) indicates $S_{i \rightarrow j}\sim \lambda_{ij}\frac{1}{t_j}$.
We repeated our analysis 
for $t_i$, i.e., the age of incumbent handset $i$ when substituted, 
finding that all curves of $p(S_{i \rightarrow j}|t_i)$ automatically collapsed onto each other (Fig.~\ref{fig:basics}F). 
Hence, when incumbents are substituted, 
whether they were released merely a few months ago (small $t_i$) 
or have existed in the market for years (large $t_i$), their substitution rates follow the same distribution, 
documenting an independence between substitution rates and the age of the incumbents. 
Mathematically, Fig.~\ref{fig:basics}F indicates $p(S_{i \rightarrow j}|t_i) = p(S_{i \rightarrow j})$.

Together, 
Figs.~\ref{fig:basics}D--F help us uncover two more mechanisms governing substitutions,
\emph{recency} and \emph{propensity}:  
substitution rates depend on the recency of substitutes, following a power law $1/t_j$.  
The uncovered power law decay has a simple origin, documenting the role of competitions in driving the obsolescence of handsets.
Indeed, when $j$ first entered the system, being the latest handset (small $t_j$),  
it substitutes for the incumbent at its highest rate.
Yet with time, more and more newer handsets are introduced. 
The constant rate of new arrivals
(Fig.~\ref{fig:basics}A) 
implies that the number of alternatives to $j$ grows linearly with $t_j$. 
Hence if we pick one handset randomly, 
the probability for handset $j$ to stand out among its competitors decays as $1/t_j$. 
The temporal decay is further modulated by the inherent propensity $\lambda_{ij}$ between two handsets, 
capturing the extent to which a certain handset is more likely to substitute for some handsets than others.
Taken together, Figs.~\ref{fig:basics}B--F predict
\begin{equation}
\Pi _{i \rightarrow j } =\lambda_{ij}N_j\frac{1}{t_{j}}.
\label{eq:Pi}
\end{equation}

\textbf{\textit{Minimal Substitution Model}}.
 Most importantly, (\ref{eq:Pi}) defines a Minimal Substitution (\emph{MS}) model, which, as we show next, naturally leads to the observed power law early growth patterns.
In this model, the system consists of a fixed number of individuals, with new handsets being introduced constantly (Fig.~\ref{fig:basics}A).
In each time step, an individual substitutes his or her current handset $i$ for new handset $j$ with probability $\Pi _{i \rightarrow j}$, according to (\ref{eq:Pi}). The propensity $\lambda_{ij}$ between handset $i$ and $j$ is
drawn randomly from a fixed distribution. 
Our results are independent of specific distributions $\lambda_{ij}$ follows.
We can solve our model analytically in its stationary state (Fig.~\ref{fig:basics}A) by plugging (\ref{eq:Pi}) into (\ref{eq:masterEq}), yielding (Supplementary Note~4):
\begin{equation}
\begin{split}
N_{i}(t_i) &= h_{i}t_{i}^{\eta_{i}}e^{- t_{i} /\tau_{i}},
\label{eq:N2}
\end{split}
\end{equation}
indicating that the number of individuals using handset $i$  is governed by three parameters: $\eta_i$, $h_i$ and $\tau_i$. 
$\eta_i \equiv \sum_k \lambda_{k\rightarrow i}N_k $ captures the \emph{fitness} of a handset, measuring the total propensity for users to switch from all other handsets to $i$.  
The \emph{anticipation} parameter $h$ arises from the boundary condition at $t_i=0$ when solving the differential equation (\ref{eq:masterEq}), 
approximating the number of individuals using handset $i$ when $t_i=1$, which  
captures users' initial excitement for a particular handset.
$\tau_i$ is the \emph{longevity} parameter, as it captures the characteristic time scale for $i$ to become obsolete.  
Indeed, defining $t^*_{i}$ as the time when a handset reaches its maximum number of users, 
equation (\ref{eq:N2}) predicts that the peak time $t^*_{i}$ is proportional to its longevity parameter and fitness: $t_{i}^*= \eta_{i} \tau_{i}$. 

The impact of handset $i$, i.e., its cumulative sales, can be calculated by integrating all transition flows from other handsets 
to $i$ before $t_i$:
$I_i(t_i) = \int_{0}^{t_i}  \sum_{k} \Pi_{k \rightarrow i} N_{k} dt$, 
yielding: 
\begin{equation}
\begin{split}
I_{i}(t_i) &= h_{i}\eta_{i} \tau_{i}^{\eta_{i}}\gamma_{\eta_{i}}(t_i/\tau_{i}), 
\label{eq:I}
\end{split}
\end{equation}
where $\gamma_{\eta}(t) \equiv \int_{0}^{t} x^{\eta-1}e^{-x}dx$
is the lower incomplete gamma function. 
Hence, in the early stage of a lifecycle (small $t_i$), 
(\ref{eq:I}) predicts that the impact of handset $i$ grows following a power law:
\begin{equation}
I_{i}(t_{i})=h_{i}t_{i}^{\eta_{i}},
\label{eq:Searly}
\end{equation}
where the growth exponent  
is uniquely determined by the fitness parameter $\eta_i$, 
equivalent to the power law exponent discovered in (\ref{eq:Itau}). 
Equation (\ref{eq:Searly}) indicates that the specific power law exponent for each constituent is  
governed by its propensity to substitute for the incumbents in the system. 
The higher the fitness, the steeper is the power law slope, hence 
the faster is the take-off in the number of users. 
The power law growth is further modulated by the anticipation parameter $h$, 
capturing the impact difference during the initial release. 
Note that it may take some time for model parameters to reach their stationary state, which may affect the validity of (\ref{eq:N2}) and (\ref{eq:Searly}). 
To this end, we performed agent-based simulations of the model, finding that the parameters reach stationary states faster than the empirical time scale we measure (Supplementary Note~4, Supplementary Note~21).

\textbf{\textit{Universal impact dynamics}}.
The \emph{MS} model not only explains the early growth phase; 
It also predicts the entire lifecycle of impacts (Supplementary Note~4).
By using the rescaled variables: $\tilde{t_{i}}=  t_i/\tau_{i}$ and $\tilde{I_{i}}=I_{i}/(h_{i}\eta_i\tau_{i}^{\eta_{i}})$, we obtain: 
\begin{equation}
\tilde{I_{i}}= \gamma_{\eta_{i}}(\tilde{t_{i}}). 
\label{eq:sg}
\end{equation} 
Therefore, for handsets with the same fitness, 
their impact dynamics can be collapsed into a single function after being rescaled by the three independent parameters ($\eta$, $\tau$ and $h$). 
Most interestingly, since the rescaling formula (\ref{eq:sg}) is independent of the particulars of a system, 
it predicts that, constituents from {\it different} systems should all follow the {\it same} curve as long as they have the same fitness.

\begin{figure} 
\centering
\resizebox{1.0\hsize}{!}
{\includegraphics{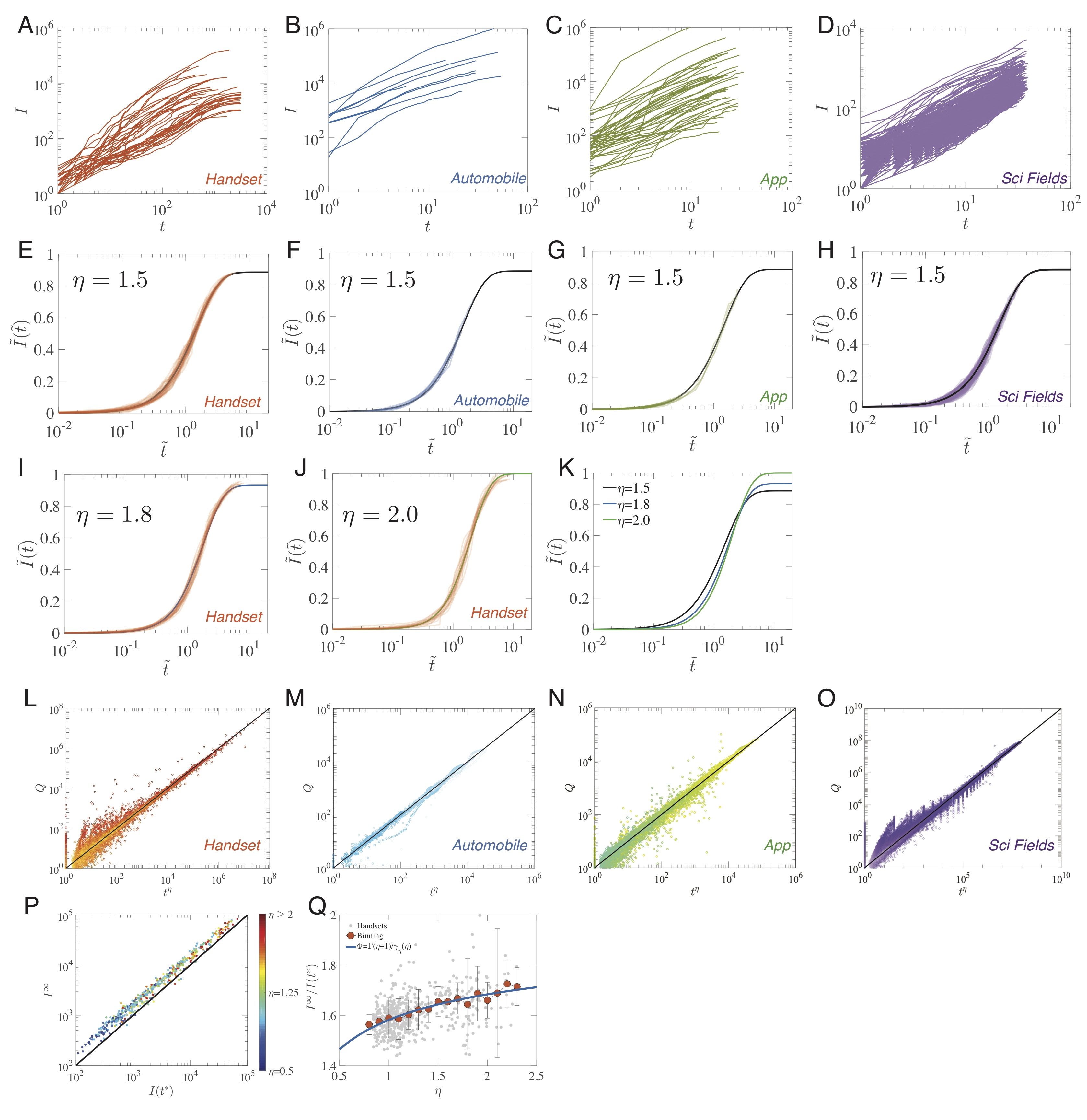}}
\caption{
} 
\label{fig:collapse}
\end{figure}

To test these predictions, 
we fit our model (\ref{eq:I}) to all four systems using maximum-likelihood estimation (Supplementary Note~4) to 
obtain the best-fitted three parameters ($\eta_{i}$, $h_i$, $\tau_i$) for each handset, automobile, smart-phone app and scientific field.  
We first selected from the four systems, those with similar fitness ($\eta \approx 1.5$).  
Although their impact dynamics appear different from each other (Fig.~\ref{fig:collapse}A--D), we find all curves simultaneously collapsed into one single curve after rescaling (Fig.~\ref{fig:collapse}E--H). 
To test for variable fitness, we selected two additional groups of handsets ($\eta \approx 1.8$ and $\eta \approx 2.0$), 
finding that the rescaled impact dynamic in both groups can be well approximated by their respective universality classes predicted by (\ref{eq:sg}) (Fig.~\ref{fig:collapse}I--J). 
The universal curves correspond to the associated classes of the incomplete gamma functions $\gamma_{\eta_{i}}(\tilde{t_{i}})$,which only depend on the fitness parameter $\eta$ (Fig.~\ref{fig:collapse}K). 
The model also predicts that if we properly normalize out the effect by $\gamma_{\eta_{i}}(\tilde{t_{i}})$, we can rescale the entire lifecycle to a power law solely governed by $\eta$. Indeed, (\ref{eq:I}) indicates that, 
by defining $Q(t) \equiv \left[I(t)/h-\tau^{\eta}\gamma_{\eta+1}(t/\tau)\right]e^{t/\tau}$, 
$Q$ should grow following a power law, $Q(t)=t^{\eta}$ (Supplementary Note~4). 
We find agreement across the four systems we studied (Figs.~4L--O).
Together Figs.~\ref{fig:collapse}A--O document regularities governing impact dynamics, which appear to hold both \emph{within} a system and \emph{across} different complex substitutive systems. 
Given the diversity of the studied systems and the numerous factors that determine the dynamics of spreading processes, 
ranging from initial seeds and timing\cite{salganik2006experimental, van2014field} to social influence\cite{aral2009distinguishing, morone2015influence} to a large set of often unobservable factors\cite{watts2011everything}, 
this level of agreement is somewhat unexpected.

\textbf{\textit{Linking short-term and long-term impacts}}.
The \emph{MS} model predicts an underlying connection between short and long-term impact. 
Indeed, we can calculate the ultimate impact---the total number of a particular handset, automobile, smart-phone app or scientific field,  
ever sold, downloaded or studied in its lifetime---by taking the $t \rightarrow \infty $ limit in (\ref{eq:I}), obtaining:  
\begin{equation}
I_{i}^{\infty}= h_{i}\Gamma(\eta_{i}+1)\tau_{i}^{\eta_{i}}, 
\label{eq:Sinf}
\end{equation}
where $\Gamma(z) \equiv \int^{\infty}_0 x^{z-1}e^{-x}dx$ corresponds to the gamma function.
Comparing (\ref{eq:I}) and (\ref{eq:Sinf}) reveals that ultimate impact and the impact at the peak number of users follow a simple scaling relationship
\begin{equation}
\frac{I_{i}^{\infty}}{I_i(t_{i}^*)}= \Phi (\eta_i),   
\label{eq:golden}
\end{equation}
where $\Phi(\eta) \equiv \frac{\Gamma(\eta)}{ \gamma_{\eta}(\eta)}$. 
That is, $I_{i}^{\infty}$ scales linearly with peak impact $I_i(t_{i}^*)$, 
and their ratio is determined only by the initial power law exponent $\eta_i$.
To validate (\ref{eq:golden}) we find  
$I_{i}^{\infty}$ and $I_i(t_{i}^*)$ follow a clear linear relationship in our dataset for different values of $\eta$ (Fig.~\ref{fig:collapse}P).
In addition, Fig.~\ref{fig:collapse}P shows the relationship posts a slight shift as $\eta$ increases. 
The rather subtle shift is also consistent with (\ref{eq:golden}), 
as $\Phi(\eta)$ increases slowly with $\eta$ (Fig.~\ref{fig:collapse}Q).  
Therefore, the uncovered power law growth patterns potentially offer a link between short-term and long-term impact in substitutive systems.

\textbf{\textit{Discussion}}.
In summary, here we analyzed a diverse set of large-scale data pertaining to substitutive processes, 
finding that 
early growth patterns in substitutive systems do not
follow the exponential growth customary in spreading phenomena.
Instead, they tend to follow power laws with non-integer exponents, 
indicating that they start with an initial explosive adoption process, 
followed by a much slower growth than expected in normal diffusion.
Analyzing patterns of 3.6M individuals substituting for different mobile handsets, we uncovered three elements governing substitutions.  
Incorporating these elements allowed us to develop a minimal model for substitutions,  
which predicts analytically the power law growth patterns observed in real systems,
and collapses growth trajectories of constituents from rather diverse systems into universal curves.

Together, 
the results reported in this paper unpack the origin of robust self-organization principles emerging in complex substitutive systems, 
and demonstrate a high degree of convergence across the systems we examined. 
Given the ubiquitous role substitutions play in a wide range of important settings, 
our results may generalize beyond the instances we studied. 
Potentially, these results could be relevant to our understanding and predictions of all spreading phenomena driven by substitutions, 
from electric cars to scientific paradigms, and from renewable energy to new healthy habits.   

This work also opens up a number of directions for future investigations.
For example, what is the role social network plays in substitutive dynamics? Unfortunately, regulations in the country from which the mobile-phone dataset was collected prohibited us from obtaining any social network information. Nevertheless, the mobile phone setting may offer a distinctive opportunity to address this question, if mobile communication records could be collected in future studies to construct social connections among users\cite{onnela2007structure,pentland2015social, eagle2009inferring, valera2015modeling, dasgupta2008social, sundsoy2010product}. 
Advances along this direction will further our understanding of substitutive dynamics and could also contribute meaningfully to the literature on social dynamics\cite{castellano2009statistical,deville2016scaling,song2010limits,gleeson2014simple, kooti2016portrait}.

Furthermore, within each system, the obtained parameters for different constituents show interesting correlations (e.g.~We find negative correlations between the anticipation parameter $h$ and fitness $\eta$, where the Pearson coefficient is -0.1642 for handsets, -0.5125 for automobiles, -0.13 for mobile applications and -0.416 for scientific fields, respectively). 
While such correlations do not affect the conclusion of the present paper, 
as our model estimates its parameters jointly and is compatible with any correlations real systems might possess (Supplementary Note~4), 
the uncovered correlations suggest interesting directions for future studies. For example, one could better understand the different forces that may affect growth patterns by collecting auxiliary information on various constituents and inspecting their correlations with the model parameters. 
Such auxiliary information could also help us better understand why diverse constituents differ from each other both within and across different systems.

On a theoretical level, it would also be interesting to explore further connections between our model with powerful theoretical tools offered by the epidemiology literature\cite{pastor2015epidemic}, such as recent findings on clustered epidemics\cite{hebert2015complex,scarpino2014epidemiological} and multi-season models of outbreaks involving multiple pathogens with different levels of immunity\cite{scarpino2016effect}. 

It is important to note that, because our model is minimal, it ignores various contextual mechanisms, such as marketing campaigns, promotional activities, or other platform-specific mechanisms, all of which could affect the studied phenomena. 
 Although we analyzed large-scale datasets from four different domains, to what degree our results can be extended beyond studied systems is a question we cannot yet answer conclusively.
However, the empirical and theoretical evidence presented in this paper provides a path toward the investigation of similar patterns in different domains, including reexaminations of familiar examples of spreading dynamics, 
as high-resolution data capturing early growth patterns become available. 
For example, there is growing evidence in the epidemiology community showing that the early spreading of certain diseases like Ebola and HIV exhibits deviations from exponential growth, featuring sub-exponential growth patterns\cite{chowell2017perspectives,viboud2016generalized, chowell2016mathematical}.  
Although power-law early growth has not received as much attention, our results suggest that 
it may be more common than we realize, and that the power law growth explained in our work may exist in even broader domains.

\newpage

\textbf{\textit{Methods}}.

{Details of studied datasets are described in the main text and Supplementary Note~1 \emph{Data Descriptions}. Empirical analyses 
of substitution patterns are detailed in Supplementary Note~2 \emph{Substitutions in Handset Dataset}. Mathematical derivations of
the minimal substitution model (Eqs.~4---8) are summarized in Supplementary Note~4 \emph{Minimal Substitution Models}.
The handset-specific parameters are obtained through maximum likelihood estimation, as described in Supplementary Note~4. The use of mobile phone datasets for research purposes was approved by the Northeastern University Institutional Review Board. Informed consent was not necessary because research was based on previously collected anonymous datasets.}

\textbf{\textit{Data availability}}.

Data necessary to reproduce the results in the manuscript are available. The automobile, smart-phone apps and scientific fields datasets are publicly available at \\ \emph{https://github.com/chingjin/substitution.github.io}. 
The mobile phone dataset is not publicly available due to commercially sensitive information contained, but are available from the corresponding author (dashun.wang@kellogg.northwestern.edu)  on reasonable requests.

\textbf{\textit{Code availability}}.
The custom codes are available at \emph{https://github.com/chingjin/substitution.github.io}. 

\newpage


%

%
%
%

\newpage

\bibliographystyle{naturemag}
\bibliography{submitbib}

\newpage
\textbf{\textit{Competing Interest}}. The authors declare that they have no competing interests.

\textbf{\textit{Acknowledgements}}.
{The authors thank B. Uzzi, J. Colyvas, J. Chu, M. Kouchaki, Q. Zhang, Z. Ma, 
and all members of Northwestern Institute on Complex Systems (NICO) for helpful comments. 
The authors are indebted to A.-L. Barab\'asi for initial collaboration on this project and invaluable feedback on the manuscript. 
This work was supported by the Air Force Office of Scientific Research under award number FA9550-15-1-0162 and FA9550-17-1-0089, 
and National Science Foundation grant SBE 1829344.
C.S. was supported by the National Science Foundation (IBSS-L-1620294) and by a Convergence Grant from the College of Arts \& Sciences, University of Miami. The funders had no role in study design, data collection and analysis, decision to publish or preparation of the manuscript.}

\textbf{\textit{Author Contributions}}.
All authors designed the research. 
C.J., C.S. and D.W. did the analytical and numerical calculations. 
C.J. C.S., J.B. and D.W. analyzed the empirical data. 
D.W. was the lead writer of the manuscript. 

\newpage
\textbf{Figure 1. Power law growth patterns in substitutive systems}.  
\textbf{(A)} Normalized impacts of all 885 handsets, which have been released for at least six months and used by 50 users in total (Supplementary Note~1). 
To compare different curves, we normalized $I(t)$ by $I(1)$, the number of users on the first day of release. 
We use the first six months to measure the early growth phase for each handset, finding that a considerable number of products do not follow exponential 
(black dotted line) or logistic growth (grey dashed line). 
Instead, they prefer power law growth patterns (statistical tests for growth comparison see Supplementary Note~1). 
\textbf{(B to D)} Similar normalized impacts of 126 automobiles (B), 2672 smart phone apps (C) and 6,399 scientific fields (D). Here we show the early growth pattern of all products whose 
records are long than their early growth period (four months for automobiles, seven days for smart phones apps and eighteen years for scientific fields), finding again that a large number of products 
prefer power law growth patten than exponential functions. 
Note that the exponential and logistic curves are shown as guide to the eye, meant to highlight a conceptual difference between exponential and power law functions. Interested readers should refer to Supplementary Figure~4, 12 and 14 for more quantitative evaluations.  
\textbf{(E to H)} (E) Normalized impacts of 240 different handsets as a function of time. 
We find, for a substantial fraction of handsets (240 handsets out of 885, 27.12\%), their early growth patterns can be well approximated by power laws ($R^{2} \geq 0.99$):
$I(t) \sim t^{\eta}$. The color of the line corresponds to the associated power law exponent for each handset, $\eta$. 
The solid lines are $y=x^{1/2}$, $y=x$, and $y=x^{2}$, respectively, as guides to the eye. 
The black dotted line corresponds to the exponential function following $y\sim e^{x}$ and the 
grey dashed line corresponds to the logistic function following $y\sim L/(1+e^{-k(x-x_0)})$,
highlighting their fundamentally different nature comparing with power law growth patterns (see Supplementary Note~1 for statistical test 
for fitting). 
\textbf{(F-H)} Similar power law growth patterns are observed in other three datasets, where we find 37 out of 126 cars (29.37\%), 1,022 out of 2,672 apps (38.25\%)
and 1,743 out of 6,399 scientific fields (27.24\%) can be well approximated by power laws. 
\textbf{(I to L)} We rescale the impact dynamics plotted in (E-H) by $t^\eta$, finding all curves collapse into $y=x$. 
 \textbf{(M to P)} Distribution of power law exponents $P(\eta)$ for curves shown in (E-H).

\textbf{Figure 2. Empirical substitution network.}
We used the backbone extraction method\cite{serrano2009extracting} to construct a substitution network, capturing substitution patterns among handsets aggregated within a six-month period (January 2014---June 2014). 
Each node corresponds to one type of handset released prior to 2014 by one of the six major manufacturers.  
Node size captures its popularity, measured by the number of users of the particular handset at the time. 
Handsets are colored based on their manufacturers (node coloring), which fade with the age of handsets. 
If users substituted handset $i$ with $j$, we add a weighted arrow pointing from $i$ to $j$.  
The link weight captures the total substitution volumes between two handsets within the six-month period.  
Since the full network is too dense to visualize, here we only show the statistically significant links as identified by the method proposed in Ref.\cite{serrano2009extracting} for p-value 0.05. 
We color the links based on the color of the substituting handset. 
The network vividly captures the widespread transitions from feature handsets to smart phones. 
Indeed, most cross-manufacturer substitution links are either yellow or green, indicating their substitutions by iPhones or Android handsets. 
Substitution patterns are also highly heterogeneous. 
A few pairs of handsets have high substitution volumes, e.g. between the successive generations of iPhones, 
but most substitutions are characterized  by rather limited volumes.   
The structural complexity shown in (A) is further coupled with a high degree of temporal variability. 
Indeed, the system turns into a widely different configuration every year, even for the most dominant handsets (Supplementary Figure~18B--E). 

\textbf{Figure 3. Empirical substitution patterns}.  
\textbf{(A)} The number of new handsets launched per quarter as a function of time. We find new handsets are introduced at a constant rate. 
The most popular handset within each eight-month time window is highlighted by an image of the handset model. 
\textbf{(B)} Substitution probability $\Pi_{i \rightarrow j}$ is proportional to $N_{j}$, consistent with the preferential attachment effect. 
Inset shows $\Pi_{i \rightarrow j}$ is largely independent of $N_{i}$. The measurement is based on eight snapshots of observations sampled uniformly in time. Specifically, we choose the first month of each year from 2007--2014 to measure substitution flows to test the preferential attachment hypothesis. 
\textbf{(C)} Distribution of substitution rates $S_{i \rightarrow j}$. Here we show the distribution as a probability density function and we measured the substitution rates among handsets in January~2014.  
\textbf{(D)} Distribution of substitution rates $S_{i \rightarrow j}$ conditional on age of the substitute $t_j$. The distributions shift systematically to the left as $t_{j}$  increases. 
\textbf{(E)} After rescaling substitution rates by $t_{j}^{-1}$, we measure $p(S_{i \rightarrow j}t_{j} | t_j)$, finding all seven curves in (D) collapse into one single curve. In the inset figure, we test the relationship between $S_{i \rightarrow j}$ and $t_j$ by rescaling $S_{i \rightarrow j}$ by $t_{j}^{-\theta}$, finding that the case where the curves collapse onto each other when $\theta=1$. 
\textbf{(F)} Distribution of substitution rates conditional on age of the incumbent ($t_{i}$). All curves collapse automatically onto one single distribution, indicating an independence between substitution rates and the age of incumbent handsets.

\textbf{Figure 4. Universal impact dynamics}. \textbf{(A to D)} Impact dynamics for products with similar fitness ($\eta=1.5 \pm 0.1$), including 40 handsets (A), 9 automobiles (B) and 43 apps (C) 
and 505 scientific fields (D). 
\textbf{(E to H)} Data collapse for products shown in (A---D). After rescaling time and impact independently by $\tilde{t_{i}}=  t_i/\tau_{i}$ and $\tilde{I_{i}}=I_{i}/(h_{i}\eta_i\tau_{i}^{\eta_{i}})$, we find all curves from four systems collapse into the same universal curve, as predicted by (\ref{eq:sg}). 
\textbf{(I)} Data collapse for handsets with similar fitness $\eta=1.8 \pm 0.1$ (30 handsets). 
\textbf{(J)} Data collapse for handsets with similar fitness $\eta=2.0 \pm 0.1$ (22 handsets).
\textbf{(K)} The universal functions shown in (E---J) are each associated with their respective universality classes that are solely determined by $\eta$. 
Here we visualize the analytical function $\tilde{I}=\gamma_{\eta}(\tilde{t})$, with $\eta=1.5$, $1.8$ and $2.0$.  
\textbf{(L to O)} The entire lifecycle can be rescaled as power laws if we properly normalize out the effect from the incomplete gamma functions. Indeed, because the function $\gamma_{\eta}(x)$ has recurrence property $\gamma_{\eta+1}(x)=\eta\gamma_{\eta}(x)-x^\eta e^{-x}$, 
(\ref{eq:I}) predicts that by defining $Q \equiv (I(t)/h-\tau^{\eta}\gamma_{\eta+1}(t/\tau))e^{t/\tau}$, we should expect $Q=t^{\eta}$.  
Here we plot $Q$ as a function of $t^{\eta}$ for all fitted products in the four systems, 
where the color of each line corresponds to the learned fitness parameter 
(See also Supplementary Note~4, Supplementary Figure~25--28 for discussions about curve collapse and comparison with other models.)  
\textbf{(P)} $I^{\infty}$ as a function of $I(t^{*})$ for the handsets with different fitness $\eta$ shown in (L).   
$I^{\infty}$ and $t^{*}$ are calculated through the system parameters: $h$, $\eta$, and $\tau$. $I(t^{*})$ is the handset's impact at time $t^*$ obtained from the empirical data (Supplementary Note~4). 
\textbf{(Q)} Scatter plot for the ratio $I^{\infty}/I(t^{*})$ as a function of $\eta$ for the same handsets shown in (P). 
The error bar indicates one standard deviation. 
The solid line corresponds to the analytical prediction by (\ref{eq:golden}). 



\section*{Supplementary material}

# Supplementary Notes

## Supplementary Note 1: Dataset Descriptions

**Dataset Introduction**

To study early growth patterns in complex substitutive systems, we explore four different domains where large-scale databases with fine temporal resolution are available:

*D1* is a mobile phone dataset recorded by a major Northern European telecommunication company. It captures daily usage patterns of 3.6 Million individuals substituting 8,928 types of mobile handsets from 01/01/2006 to 11/03/2014. By identifying each anonymized SIM-ID as an individual user, detecting the first and the last date when the individual used a particular handset, we construct the usage timeline for every handset model. To measure the impact of each handset $I(t)$, we calculated the number of individuals who bought the handset up to time $t$ since its availability. Here, we specifically focus on 885 handset models in the dataset to study the early growth pattern of handset impacts. These handset models were chosen because they have been released for at least 180 days and have at least 50 users in total to make sure we have enough statistics for our data analysis and to be able to observe the early growth period (Definition of the early growth phase, see *Identifying the Early Stage of the Growth Curve* and Supplementary Figure 1).

*D2* is an automobile dataset collected from a website that records automobile sales data

(*Good Car Bad Car*). The dataset captures monthly transaction records of 135 different models of automobiles sold in the U.S. and Canada from year 2010 to 2016. We focused on 126 models that have been introduced to market for more than 4 months to guarantee we have enough data to study their early growth patterns. In this dataset, we define the impact of automobiles as their cumulative sales across North America.

*D3*, a smartphone application dataset, captures daily-download records for the top 2,672 mobile apps released between 11/20/2016 and 12/20/2016 in the App store by Apple. Here we only focus on apps within one single operating system: the iOS system. These apps have been introduced for at least two weeks, allowing us to study their early growth patterns. In this dataset, the impact of each app is defined as the cumulative number of downloads. The data are collected from a mobile application platform *Apptopia: https://www.apptopia.com/*. The website collects information for each mobile app and categorizes them based on their functions. For each category, the website also ranks the apps by their performance, and selects for the top apps released within the latest month.

*D3s. Health and Fitness Apps*. To avoid possible selection bias towards highly popular apps that may skew our empirical observations, we compiled another complementary dataset to form a more uniform sample, consisting of all 70,377 iTunes applications belonging to the category *Health and fitness* until 12/15/2016 from *Apptopia*. Since the information is most complete within 3 months, we studied 22,982 apps released after 9/15/2016. We repeated the same analysis as Fig. 1 on this dataset, uncovering same power law growth patterns (Supplementary Figure 2).

*D4*, a scientific field dataset, which is obtained from the Microsoft Academic Graph. Different from Google scholar, MAG specializes on semantic search, hence has an excellent entity resolution engine. As such, it offers the state of art classification of scientific fields. We curated this entirely new dataset, including 172,037,947 publication records for 209,404,413 scientists for more than one century. Linking the publication records to 228,563 scientific topics provides us the largest dataset of the four substitutive systems, allowing us to analyze substitution behavior among scientific topics. More specifically, we studied 246,630 scientists who are active from 1980 to 1990, substituting for 6,399 fields of studies which are initiated between 1980 and 1990. We trace the impact dynamic of the fields from 1980 to 2018. By measuring the number of scientists who have published papers in the field, we are able to quantity the early growth pattern like we did for all other datasets. All studied fields have been studied for at least 28 years which is long enough for us to explore the early growth pattern for each field.

**Identifying the Early Stage of the Growth Curve**

To focus on early growth patterns, we systematically define the concept *early growth phase* in four studied datasets to make sure that we are consistent across our analysis. Specifically, for each item in a given system, we identify the time $T_s$ when the growth rate $dI/dt$ reaches its peak ($d^2I/dt^2 = 0$). For each dataset, we define $T^*$ as the position of the first highest peak of the distribution of $T_s$ (Supplementary Figure 1A–D) and the period $t \leq T^*$ as its *early growth phase*. We find $T^* = 180$ days for handsets, $T^* = 4$ months for automobiles, $T^* = 7$ days for apps and $T^* = 18$ years for scientific fields. The differences in $T^*$ across the four systems agree with

our intuition of the typical lifecycle differences among handsets, automobiles, mobile apps and scientific fields. Note that the early growth period for cars and handsets are shorter than the typical lease or loan duration for such products. Hence, the observed growth patterns are unlikely to be affected by these factors.

**Early Growth Pattern**

In Supplementary Figure 1E–L, we show the early growth pattern for different products in Fig. 1. Instead of normalizing the impacts with $I(1)$ (the number of users on the released date), we show the original impact dynamics for the products, finding they tend to follow power law growth patterns. Interestingly, we discover that handsets and automobiles with high $I(1)$ are not associated with high power law exponent $\eta$. To systematically study this phenomenon, we plot the relationship between $I(1)$ and $\eta$ in Supplementary Figure 1M–P, finding the phenomenon is system-dependent. While $I(1)$ and $\eta$ are moderately negative-correlated for handsets and automobiles, suggesting products that registered the most sales tend to have a slower build up in its sales, the two parameters are largely independent and weakly correlated of each other for smartphone apps and scientific fields. This indicates that it is possible that $I(1)$ and $\eta$ are not driven by the same mechanism, pertaining to different processes governing substitution dynamics. Therefore, in our modeling framework, we do not assume any specific correlations between the power law exponent $\eta$ and $I(1)$, allowing any possible relationship between the parameters. They could be either uncorrelated, positively/negatively correlated, depending on different systems, which are all compatible with our modeling framework. Notice that the number of products shown in Supple-

mentary Figure 1 represents a rather substantial fraction of visible products within the system given that impact typically follows fat-tail distributions (Supplementary Figure 3). Indeed, most products have relatively small impacts, limiting the number of samples to study their impact dynamics properly.

To systematically study how well the early growth patterns could be fitted as power laws, we compare the power law fit with alternative function forms. We first test exponential growth pattern by plotting the rescaled impact dynamic in semi-log plot (Supplementary Figure 4A–D). If the early growth pattern initially follows an exponential growth, we expect to observe a straight line in the early stage. As shown in Supplementary Figure 4, the growth curves resemble closely what power laws would look like on a semi-log plot, showing clear deviations from an exponential function. To systematically compare the exponential and power law fits, we apply two different statistical tests: 1) R-square (Supplementary Figure 4E–H) and 2) Akaike Information Criterion test (Supplementary Figure 4I–L). The two tests quantify the performance of the two models in the overall fitting samples, showing the power law fit clearly outperforms the exponential fit. Specifically, the AIC scores show that 99.21% handsets, 93.69% automobiles, 92.43% smartphone apps and 80.18% scientific fields prefer a power law to an exponential fit. Furthermore, since our focus is on early growth, to test specifically whether power law provides a better fit in small $t$ region, we adopt a third statistical method: weighted Kolmogorov-Smirnov (KS) test[1], measuring the performance of each product through

$$D_i = max_{t \in [0,T]} \frac{|I_i^t - \tilde{I}_i^t|}{\sqrt{(1 + I_i^t)(I_i^T - I_i^t + 1)}}. \tag{1}$$

We find the weighted KS test provides a normalized measure of the goodness of fit for different

5

stages. The power law function again outperforms an exponential fit in all four systems (Supplementary Figure 4M-P). It is also worthy to note that, while the majority of products follow power law growth patterns, for some of the cases in each of the four systems, their impacts start to saturate at a rather early stage, sometimes earlier than $T^*$ (Supplementary Figure 5), corresponding to cases with a low $R^2$, as shown in Supplementary Figure 4E–H. We also find, in soft substitution systems such as mobile applications or scientific fields datasets, a minority of growth patterns indeed prefer an exponential growth to power law growth patterns, indicating that traditional spreading process could be coexistent with substitutive process in such systems. In Supplementary Figure 6, we also show the rescaled early impact dynamics for all items in the four datasets, as we have done in Fig. 1. We find that, although as expected, growth patterns of all products show more variations around $y = x$, they exhibit clear difference from exponential growth patterns.

Could the growth pattern be explained by other alternative models, such as linear function? To answer this question, we systematically study the 95% confidence interval of the fitted value as a function of $\eta$ (Supplementary Figure 7A-H). We find the confidence interval increases with the fitted exponent $\eta$. This raises the possibility that some curves may be fitted as $\eta \neq 1$, but are nevertheless within the confidence interval of $\eta = 1$. To test this, we measure the fraction of exponents whose confidence interval touches $\eta = 1$. We first measured these fractions within the curves with high fitting performance ($R^2 > 0.99$), finding that only a small fraction of products may be approximated by linear growth (Supplementary Figure 7A-D Handsets 7.9%, Automobiles 8.1%, Apps 5.77% and Scientific Fields 6.02%). We then relaxed the fitting performance, finding that, even in a more generous case, the vast majority of products lie outside the confidence interval

where $\eta = 1$ might be a viable candidate exponent (Supplementary Figure 7E-H). We further use AIC score to systematically compare the power law growth and linear growth model by controlling for the number of parameters of a model. AIC is defined as $AIC = 2k - 2\log(\max(L))$, where $k$ corresponds to number of parameters in the system, and $L$ represents the maximum value of the likelihood function of the model. When evaluated by this measure, the preferred model should yield a lower AIC comparing with its competitor. Therefore, we fit each product with the two models, selecting the better fitted one with the lower AIC. We find, only a very small proportion of products (4.25%-8.75%) prefers linear growth model (Supplementary Figure 8). A vast majority of the products prefer a power law growth model than a linear growth pattern.

### Alternative Definitions of Early Growth Phase

Are these results remain robust if we take alternative definitions of early growth phase? Here, we first test the robustness of these results with two alternative definitions of $T^*$. 1) The mean of $T_s$: We test our results by taking the mean value of $T_s$ in each system to estimate the early growth pattern, where the early growth period is defined as $[0, T^*]$ ($T^* = 280.4$ days for handsets, 19.45 months for cars, 9.33 days for mobile apps, 18.4 years for scientific fields which remains the same). We find the results are robust to this new definition, where only a very small fraction of products (3.6%-8.44%) can be explained by linear growth models (Supplementary Figure 9A-C). 2) The median: We repeated our analysis by taking the median value of $T_s$ to estimate the early growth pattern ($T^* = 221$ days for handsets, 15 months for cars, 7 days for mobile apps and 18 years for scientific fields which remains the same). We find again, only a very small

7

fraction of products (5.77%-9.95%) can be explained by linear growth models (Supplementary Figure 9D-F). This finding remains the same even we extend the data to all products with long enough records (Supplementary Figure 10). To further test the robustness of our estimation of the power law exponents, we compare $\eta_{mode}$, $\eta_{mean}$ and $\eta_{median}$, the different power law exponents obtained by taking the mode, the mean and the median of the distribution of $T_s$ in defining the early growth period. We find the ratio of the fitted value remains relatively stable for different values of exponents across all datasets (Supplementary Figure 11). Note that, for scientific fields, the median value and mean value remain unchanged to the mode of $T_s$, therefore the results remain the same.

Next, we use AIC score to further compare the power law growth pattern to three alternative functions together (linear, exponential, logistic), finding again that for a majority of products, power law early growth pattern provides the best fit (Supplementary Figure 12, 93.67% of handsets, 81.51% of automobiles, 74.59% of mobile apps and 71.79% of scientific fields). Since the linear function also belongs to power law growth, we have 98.6% handsets, 83.5% automobiles, 79.6% apps and 74.1% scientific fields prefer power law growth pattern, rather than exponential-class functions.

Until now, we have already tested three possible definitions of early growth period by locating the mode, mean and median of the $T_s$ distribution. However, we do not know whether these definitions are robust among products with different time scale. For example, is that possible a logistic curve with a very long time scale, behaves similar to power law functions at the defined

8

early growth period? To test the robustness of the definition, here we perform a simple experiment. We consider 500 logistic growth curves, where their three parameters are randomly selected from different parameter regions (See Supplementary Table. 2), $I^\infty \in [10^0, 10^6]$, $k \in [0.03, 0.07]$ and $t_0 \in [80, 140]$. Now, we can use the same definitions of early growth period to identify the early growth trajectory for each curve. Our hypothesis is, if the definitions are robust to different time scales, our previous method should be able to classify these early growth curves as "logistic functions" instead of other function types. To test this hypothesis, we measure the inflection point $T_s$ for each curve (Supplementary Figure 13A), finding that the distribution of $T_s$ shows a similar shape as what we observed in the real datasets. $T_s$ is somewhere within the range of 70 to 140 (Supplementary Figure 13B), indicating that the typical time scales differ across different curves. We define the early growth phase of the curves as $t \leq T^*$, where $T^* = mean(T_s)$, we then fit the curves with the four functions as presented in Supplementary Table. S2, finding that 499 out of 500 curves are classified as logistic growth. Only one curve has been classified as exponential growth, and none of the curves can be regarded as power law growth or linear growth patterns (Supplementary Figure 13C). We also test alternative definition of $T^*$ by using the mode of the distribution ($T^* = 108$, the mode is equal to the median in this case), finding the result remains unchanged (Supplementary Figure 13D). This experiment confirms the robustness of the definition of the early growth phase. If the growth pattern observed in the four datasets favors a logistic growth instead of a power law, it will be classified to logistic growth (or exponential function) directly just as what we have seen in Supplementary Figure 13.

Although this experiment indicates that our previous method is somewhat robust to the defi-

nition of the early growth phase, we further test one more alternative definition of $T^*$ which allows variability of individual time scales. Here we consider a dynamical definition of the early growth phase by defining $T^* = T_s$ for each product in the four systems. Since $T_s$ is different for each individual product, the new definition allows variability of early growth phases for products with different temporal dynamics. In Supplementary Figure 14, we repeated our analysis by fitting the growth patterns to four different functions, finding a vast majority of products/fields again prefer power law growth patterns (90.35% of handsets, 80% of automobiles, 76.9% of mobile apps and 72.63% of scientific fields), documenting again the robustness of the observed power law growth patterns.

**Test the Robustness of Fitting**

To further validate the robustness of our method, we perform two levels of analysis. We first test our fitting method on a synthetic exponential system (Supplementary Figure 15A), where we know as a ground truth that these curves must follow exponential growth. Hence, if we use our method to fit them, we should recover the exponential patterns rather than power law. This is also a way to make sure that our method would not over-fit. Specifically, we generated 100 exponential curves with different exponents, and use our method to identify them. We find, all of them are correctly classified as exponential instead of power law, demonstrating our method is reliable in this case (Supplementary Figure 15A inset).

Next, we seek out real datasets capturing a system that is unlikely to be driven by substitutions. Here we use a publicly available dataset about flu spreading in US from 2003-2018. We

downloaded data from the website of Centers for Disease Control and Prevention (CDC), where the number of infections for 11 Emerging Infections Program (EIP) states in USA have been recorded in each influenza season from 2003 to 2018.

To demonstrate our fitting procedure, in Supplementary Figure 15B, we show the fitting of our method to five selected early growth dynamics. We find that these growth patterns are well approximated by straight lines in a log-linear plot, demonstrating that they prefer exponential fit to power laws. (We fit each curve with power law and exponential separately, and measure difference of Akaike Information Criterion (AIC): $\Delta AIC = AIC(powerlaw) - AIC(exponential)$. If it is positive, the curve prefers exponential than power law.)

We then apply our method systematically to all 168 curves (Supplementary Figure 15C) with enough data points (at least 5 non-zero data points), finding that a vast majority of cases (89.31%) indeed prefer exponential early growth pattern to power laws (Supplementary Figure 15C inset). We evaluated our fitting results using both *AIC* and R-square, finding consistent results for both cases. All of these analysis indicate the robustness of the method.

**Singularity of Power Law with Non-integer Exponents**

Power law with non-integer early growth pattern is rather unexpected, because it is a non-analytic function, which is a rare form to find in the case of spreading processes. Indeed, most of growth patterns observed in nature follow analytical growth patterns. Take the epidemiological models as an example, which normally assume that disease spreads through a multiplica-

tive process: the disease starts from an initial seed and infects other people with a constant rate. Mathematically, the early growth pattern of the disease is described by a differential equation: $dI(t)/dt = bI(t)$, which predicts that the early growth pattern follows an exponential growth (Supplementary Table. 1). Any analytical functions (including the exponential function) can be expanded as a Taylor series. More specifically, for analytic function $f$, we may expand it around 0: $f(x) = \sum_{n=0}^{\infty} \frac{f^{(n)}(0)}{n!} x^n$, where $f^n(x)$ is the $n^{th}$ derivative of $f$. However, if the power law growth function has a non-integer $\eta$, the $n^{th}$ derivative of $f$ would diverge if $n > \eta$ (Supplementary Table. 1). Such singularity suggests a fundamental different mechanism may be at work. We will discuss more about spreading dynamics and relevant models in Supplementary Note 3.

## Supplementary Note 2: Substitutions in Handset Dataset

One important common characteristic among four studied systems discussed in Supplementary Note 1 is that they evolve by substitutions. Although there has been a profusion of empirical studies with the recent big data explosion, particularly those emerging from online domains, tracing and measuring substitution patterns empirically have remained as a difficult, often elusive task. This may seem puzzling given the fact that models that can be used to describe substitution processes have existed for over a century[2–7]. Here we explain this situation by highlighting the key challenges that have long prevented researchers from empirical studies of substitutions, and how mobile phone datasets used in our study offer a unique opportunity to allow us to present among the first empirical evidence on substitutions.

**Challenges in Empirically Studying Substitutions**

The lack of empirical knowledge about substitution patterns is rooted in the significant, systematic challenges in collecting adequate datasets to empirically trace and measure substitution patterns:

Challenge One (*C1*): Substitutions depend strongly on time, often signaling the beginning and end of a lifecycle. Hence measuring substitutions requires longitudinal datasets that can cover a longer time period than a typical lifecycle, rendering obsolete many datasets, particularly those emerging from online settings, which span comparably or less than the typical lifetime[8,9].

Challenge Two (*C2*): Substitution implies a competitive process, in which we choose one or few out of many alternatives to substitute for. Therefore, understanding substitutions requires us to observe both the substituted ones and the alternatives. Yet studies that are potentially relevant typically involve a single[5,10,11] or an incomplete set[6,12] of substitutes, hence inevitably focus on the substituted ones, by implicitly ignoring the alternatives. This is further confounded by the well-known heterogeneity in complex systems as popularity follows a fat-tailed distribution[13–17].

Challenge Three (*C3*): Substitutions involve both substitutes and the incumbents. To observe substitutions we need to go beyond aggregated records to obtain individual level substitution histories. Otherwise, even in cases where datasets (occasionally) met *C1* and/or *C2*, it is nearly impossible to infer accurately which substitutes for which[18,19].

Here we take advantage of the increasing availability of rich databases in a ubiquitous setting, allowing us to systematically alleviate and combat all three aforementioned challenges: mobile telephony in the telecommunication sector. Indeed, mobile phones have existed with high penetration in developed countries for over a decade. Since the average usage time of a phone is less than two years, it offers an observation window that far exceeds the typical life cycle of the substitutes, in doing so eliminating *C1*. Carriers for billing purposes monitor all handsets that have ever operated within the network, ensuring the completeness in the set of substitutes we study (*C2*). Anonymized phone numbers together with their portability across devices provide individual traces for adoption and discontinuance histories, offering an excellent proxy of substitutions at an individual level within a societal-scale population, hence resolving *C3*.

**Substitution Patterns in Handset Dataset**

We start by analyzing the macroscopic properties of the mobile phone dataset and measure the total number of active handsets/users in the system as a function of time (Supplementary Figure 16). We find both quantities saturate to a constant $N = 2.5 \times 10^6$, indicating that the system reaches to a dynamical equilibrium around 2011. It also suggests that each individual in the dataset is holding one single product on average at a time. enabling us to compile the substitution timeline for each user accordingly (see Supplementary Figure 17A for an illustration) and generate a dynamic network characterizing substitution patterns among handsets.

In order to uncover the basic properties of this substitutive system, we specifically focus on an aggregated network capturing substitution patterns among 558 handsets within a six-month period

14

01/01/2014 — 06/01/2014, of which we have shown the backbone in Fig. 2. While the network has a large average degree ($\langle K \rangle = 73.6$), suggesting handsets are substituted by a considerable number of other handsets, the in (out)-degree distribution of the network follows a fat-tailed distribution (Supplementary Figure 17B), indicating a high heterogeneity in substitution selections. We also measured the distribution of substitution flows between two handsets, represented by the weight of the links, finding the distribution also follows a fat-tailed distribution (Supplementary Figure 17C). In addition to the structural complexity depicted in Supplementary Figure 17B–C, substitution patterns are characterized by a high degree of temporal variability. Indeed, the system turns into widely different configurations every year (Supplementary Figure 18B–E), driving the rise and fall patterns of handset popularities (Supplementary Figure 18A),

## Supplementary Note 3: Existing Models

Over the past century, a considerable number of studies have been devoted to understanding spreading and contagion processes from a wide range of fields: from economics and sociology[20–25] to computational social science[26–29], from epidemiology[30,31] to computer science and physics[32–39], giving birth to an immense number of mathematical models.

In this section, we classify the existing models into five different categories. For each of the category, we select among the most relevant models to show their analytical solutions and demonstrate why *none* of the them can explain the power law growth patterns observed in our

15

data.

We will also discuss the relationship between a few selected models and the Minimal Substitution model (*MS* model) proposed in our paper. In Supplementary Table. 3 we summarize for several existing models their analytical solutions and early behaviors.

**Diffusion of Innovations Models**

(**Logistic Model**) The logistic model (also known as the *SI* model in epidemiology) is widely utilized to model population growth, product adoption[40] and epidemic spreading[31], with application in many fields. In the context of production adoption, people from a conservative system are categorized as two different types: potential users and current users. In each time step, potential users are affected by current users to adopt the product with a certain probability $q$. With time, the attractiveness of the product decays, as the product have been adopted by all potential users in the system, the number of current users approaches a constant $I^\infty$, capturing the ultimate impact of the product. This process can be expressed in a rate equation:

$$\frac{dI_i}{dt} = q_i I_i (1 - I_i/I_i^\infty), \qquad (2)$$

yielding

$$I_i(t) = \frac{I_i^\infty}{1 + e^{-q_i(t-\tau_i)}}, \qquad (3)$$

where $I_i^\infty$, $q_i$ and $\tau_i$ capture the ultimate impact, longevity, and immediacy of a product, respectively. By taking $t \to 0$, we obtain the early growth pattern predicted by the model, corresponding

16

to an exponential growth pattern:

$$I_i(t)|_{t\to 0} = I_i(0)e^{q_i t}, \tag{4}$$

where

$$I_i(0) = \frac{I_i^\infty}{1+e^{q_i \tau_i}}, \tag{5}$$

captures the number of initial users of the product.

(**Bass Model**) First proposed by Frank Bass in 1969, the Bass model[41,42] is widely used in marketing, management science and technology forecasting. It describes the process through which new product are adopted by mass populations. The Bass model classifies the adopter into two groups: innovators who are mainly influenced by the mass media and imitators who adopted the product through the *word of mouth* effect. Mathematically, this can be expressed as

$$\frac{dI_i}{dt} = (p_i + q_i I_i/I_i^\infty)(I_i^\infty - I_i), \tag{6}$$

where the impact of a product $I$ is defined as the number of users. $p$ describes the probability for innovators to adopt the product, reflecting a social influence effect that is independent of the current product impact. $q$ captures the imitation process, where potential users are influenced by previous users with probability $q$. $I^\infty$ defines the ultimate impact of the product, capturing total number of users of the product. Solving the model yields

$$I_i(t) = I_i^\infty \frac{1-e^{-(p_i+q_i)t}}{1+\frac{q_i}{p_i}e^{-(p_i+q_i)t}}. \tag{7}$$

By taking $t \to 0$, we obtain the early growth pattern of the model,

$$I_i(t)|_{t\to 0} = I_i^\infty p_i t \tag{8}$$

17

which corresponds to a linear growth pattern, different from the non-integer power law growth observed in our data.

(**Gompertz Model**) The Gompertz model, named after Benjamin Gompertz, was first proposed to model mortality[43]. It has also been widely adopted to model market impact and product penetration[44,45]. The model can be formulated as

$$\frac{dI_i}{dt} = q_i I_i ln(I_i^\infty/I_i). \tag{9}$$

Solving the equation, we have:

$$I_i(t) = I_i^\infty e^{-e^{-(a_i+q_i t)}}. \tag{10}$$

The equation predicts that the product initiates from a finite number of users $I_i(0) = I_i^\infty e^{-e^{-a_i}}$, and grows exponentially at early stage:

$$I_i(t)|_{t \to 0} = I_i(0) e^{e^{-a_i} q_i t}. \tag{11}$$

**Substitution Models**

(**Fisher-Pry Model**) The Fisher-Pry Model is considered as one of the earliest substitution models[5]. It has been applied to model different substitution processes, from Synthetic/Natural Rubbers to Plastic/Natural Leathers. The model focuses on a two-product system, describing how a new product substitutes for an old one. Since only two products are considered, the model can be considered as mathematically similar to the logistic model, predicting a logistic growth pattern of the new product. Therefore, the early growth pattern predicted by Fisher-Pry model is exponential as well.

(**Lotka-Volterra Competition Model**) The Lotka-Volterra Competition (LVC) model, is frequently used to model population dynamics in biological systems. Along with its many variants, the model is widely applied to describe interaction dynamics: from species interactions to parasitic and symbiotic relations to technology competitions[3,4,46–48].

Here, we study the original version of the LVC model for a two-competitor system. Note that the model can be easily generated to a multi-product system, but the original LVC model is sufficient to illustrate the early behavior of products. The model contains two non-linear differential equations, capturing the population dynamics of the system:

$$\begin{aligned}\frac{dN_i}{dt} &= \frac{q_1 N_i}{K_i}(K_i - N_i - \alpha_2 N_j) \\ \frac{dN_j}{dt} &= \frac{q_2 N_j}{K_j}(K_j - N_j - \alpha_1 N_i).\end{aligned} \quad (12)$$

In (12), we denote $N_i$ and $N_j$ as the number of current users of the incumbents and substitutes. The competition between products are captured by coupling terms in both equations and controlled by the positive coefficients $\alpha_1$ and $\alpha_2$. Note that $\alpha_1$ and $\alpha_2$ do not necessarily equal to each other, indicating that the influences of the two products on each other can be different. $K_i$ and $K_j$ capture the market size of each technology, equivalent to their ultimate impacts in the absence of competition.

To obtain the early growth pattern of an entrant, we assume that the incumbent dominants the market when the new entrant is introduced. We studied the asymptotic temporal behavior of $N_j$ around the fixed point $(N_i = K_i, N_j = 0)$, obtaining:

$$\frac{dN_j}{dt} = q_2 N_j - \frac{q_2 N_j}{K_j}\alpha_1 K_i, \quad (13)$$

19

where the $q_2 N_j$ term corresponds to an exponential growth of the substitute, and the $\frac{q_2 N_j}{K_j} \alpha_1 K_i$ term reflects the discontinuance of $j$ due to the competition with $i$. By solving (13), we obtain the early growth pattern of the substitutes:

$$N_j(t)|_{t \to 0} = N_j(0) e^{q_2(1 - \frac{\alpha_1 K_i}{K_j})t}, \tag{14}$$

which is an exponential function. We can also attain another exponential growth of $j$'s impact by solving $\frac{dI_j}{dt} = q_2 N_j$:

$$I_j(t)|_{t \to 0} = I_j(0) e^{q_2(1 - \frac{\alpha_1 K_i}{K_j})t}. \tag{15}$$

(**Norton-Bass Model**) The Norton-Bass (NB) model was proposed by Norton and Bass in 1987 aiming at describing multi-generation diffusion processes[6,42]. Inspired by the seminal Bass model[41], the NB model consider the penetration of technology that evolves rapidly in successive generations.

The NB model consists of $k$ nonlinear equations describing the sales of $k$-generation technologies with continuous repeat purchasing. For simplicity, here we consider a system of two generations, a more complex $k$-generation case can be generalized in a straightforward manner from the following results. According to the NB model, we have:

$$\begin{aligned} N_i &= K_i F_i(t_i) - K_i F_i(t_i) F_j(t_j) \\ N_j &= K_j F_j(t_j) + K_i F_i(t_i) F_j(t_j), \end{aligned} \tag{16}$$

where $t_i$ and $t_j$ represent the age of the old generation production ($i$) and the new product ($j$). $K_i$ and $K_j$ capture the market capacity of the products. $N_i$ and $N_j$ measure the product sales. Notice

that the original NB model is designed for the product with continuous, repeated purchases, the sales at a given time can be approximated as the current number of users of a given product. The function $F_g(t_g)$ takes the following form:

$$F_g = \frac{1 - e^{-(p_g+q_g)t_g}}{1 + \frac{q_g}{p_g}e^{-(p_g+q_g)t_g}}, \qquad (17)$$

which is derived from the Bass model (see Eq. 6 and Eq. 7). The interaction between products is captured by the coupling terms $K_i F_i(t_i) F_j(t_j)$, without which the behavior of the products follows the original Bass model. To understand the behavior of the NB model, we take the limit $t \to \infty$, obtaining $N_i = 0$ and $N_j = K_i + K_j$, which indicates that the new product will take over the entire market.

From (16), we derive the asymptotic temporal behavior of $j$ around $t_j \to 0$, yielding:

$$N_j(t_j)|_{t_j \to 0} = p_j(K_j + K_i \frac{1 - e^{-(p_i+q_i)t_\Delta}}{1 + \frac{q_i}{p_i}e^{-(p_i+q_i)t_\Delta}}) t_j, \qquad (18)$$

where $t_\Delta = t_i - t_j$ measures the age difference of the products. Eq. 18 indicates that the early impact dynamics of $j$ can be approximated by linear growth patterns, hence different from the non-integer power law growth observed in our data.

**Epidemic Models**

(**SIR Model**) Epidemic models are another class of models that can be generalized to describe substitutions. One of the most famous models in this class is the SIR model[30, 31, 49]. Here people are classified into three groups: $S$ represents the susceptibles, measuring the number of people who are susceptible to adopt a product; $I$, the infectious, measures the number of people

21

who currently use the product; and $R$, the recovered group captures people who had bought the product previously, but have discontinued using it. In each time step, a current user "infects" a susceptible user with probability $\beta$, and at the same time, the user may abandon the product (recover) with certain probability $\gamma$. To avoid confusion over $I$ as impact throughout the paper, here we use $A$ to represent the number of current users, corresponding to the quantity that is typically described as $I$ in the SIR model. Mathematically, the model could be expressed as a set of ordinary differential equations,

$$\begin{aligned}\frac{dS}{dt} &= -\frac{\beta AS}{N_0} \\ \frac{dA}{dt} &= \frac{\beta AS}{N_0} - \gamma A \\ \frac{dR}{dt} &= \gamma A\end{aligned} \tag{19}$$

where $N_0$ captures the total number of people in the system, and the impact of the product can be obtained through its definition: $I(t) \equiv A(t) + R(t)$. The model does not have a closed form solution, but we can approximate the earlier behavior of the growth pattern analytically, finding that it follows an exponential growth at $t \to 0$:

$$A(t)|_{t \to 0} = A(0)e^{(\beta - \gamma)t}. \tag{20}$$

We can also derive the early growth pattern of $I$, which also follows exponential growth:

$$I(t)|_{t \to 0} = I(0)e^{(\beta - \gamma)t} \tag{21}$$

In fact, although the entire dynamic of other epidemic models (SIS, SIRS) are different from the SIR model, their early growth patterns follow the same exponential growth pattern. For a more comprehensive review of this body of literatures, refer to Ref.[49].

(**Multiple Epidemics Model**) Multiple Epidemics model was previously proposed to describe competitions among diseases. This type of models has been generalized to understand scientific paradigms shifting[50,51]. The original model focuses on a two-dimensional lattice where each site represents a particular user. In each time step, one attempts the following two moves: 1) A random site $i$ is selected and $i$ will randomly choose one of its four neighbors $j$. If $i$ has not used $j$'s current product before, she/he would adopt the product; Otherwise, the system remains the same. 2) With probability $\alpha$, another random site $k$ is selected and a newly introduced product will be assigned to the node occupying site $k$.

Monte Carlo simulation shows that in this model, the early growth pattern of product impact is determined by the lattice dimension. The model predicts that a product's impact growth rate increases linearly at its early stage in a two-dimensional lattice, indicating impact follows a quadratic growth pattern at beginning.

If we change the lattice assumption to a random graph the model predicts that the growth pattern follows an exponential growth. Therefore the class of models lacks the mechanisms to explain the divergent behavior in small $t$ region predicted by the non-integer power law exponents.

(**Sub-exponential Growth Model**) While most epidemic models predict exponential growth, recent sub-national epidemiological data at the level of counties or districts offered new observations that infectious diseases spreading via close contacts (sexually-transmitted infectious diseases, smallpox, and Ebola) exhibit sub-exponential early growth patterns[52–54]. As pointed out[53,55–57], the

23

observed sub-exponential early growth patterns are consistent with the formalism:

$$\frac{I_i(t)}{dt} = r_i I_i(t)^{p_i}, \tag{22}$$

where $r$ captures the growth rate of the disease and $p$ is the "deceleration of growth" parameter. When $p = 0$, a linear growth pattern is expected, whereas $p = 1$ would generate an exponential growth pattern. Models with similar forms have also been applied to describe innovation diffusions (see review[18]). Early growth patterns start to attract some attention in the epidemiology community as well. In particular, a recent review paper[58] and relevant comments that followed [59–63] discussed growing evidence that shows the early spreading of certain diseases like Ebola and HIV exhibits deviations from exponential growth, featuring sub-exponential growth patterns. While various hypotheses that may be responsible for the sub-exponential growth are discussed, lacking detailed datasets tracing the early spreading patterns, it has been understandably difficult to uncover the mechanisms. One key contribution of our work is to offer a mechanistic explanation for the observed power-law early growth, based on empirically falsifiable assumptions that were mined directly from large datasets. While the mechanistic explanation for sub-exponential growth in the epidemic context remains missing, these examples suggest that the power law early growth patterns we observed in our paper may possibly extend to broader domains.

**Network Growth Models**

Network growth models represent a well-known branch of models that are often associated with power laws[64–67]. Next, we will first discuss two types of network growth models: 1) Evolving network models that explain degree dynamics and heterogeneity, such as the BA model[17] and the

fitness model[64,65]; 2) Network densification models that explain the growth in the number of nodes and links[66,67]. We will then demonstrate that power laws generated by network models vis-a-vis what is observed in substitutive systems pertain to fundamentally different processes.

(**Evolving Network Models**) The fitness model (also known as Bianconi-Barabási model) was proposed to model the evolution of a competitive networked system[64,65]. At each time step, new products (represented by nodes) are introduced at a constant rate. They link with existing nodes with probability

$$\Pi_i \propto \eta_i I_i(t), \qquad (23)$$

where fitness parameter $\eta_i$ quantifies the likelihood of product $i$ to be adopted by users, $I_i(t)$ corresponds to the product impact, i.e., the degree of node $i$, capturing the well-known preferential attachment mechanism. If we set $\eta_i = 1$ for all nodes $i$, the model reduces to the BA model[17].

The fitness model predicts that, the node dynamics follow a power law growth, with the exponent governed by fitness:

$$I_i \propto t^{\frac{\eta_i}{C}}. \qquad (24)$$

$C$ is a global parameter in the interval $(\eta_{max}, 2\eta_{max}]$, which can be obtained from the following equation:

$$1 = \int_0^{\eta_{max}} d\eta P(\eta) \frac{1}{\frac{C}{\eta} - 1}, \qquad (25)$$

where $P(\eta)$ is the distribution of $\eta$. Therefore, the fitness model can predict a power law growth, but only for exponents that are in the interval $[0.5, 1)$. Notice that $\eta = 0.5$ corresponds to

25

the prediction of the BA model, which can be treated as a special case of the fitness model in this regard[17].

(**Network Densification Models**) The seminal work by Leskovec, Kleinberg and Faloutsos[66] outlines another mechanism for power law growth to emerge in the network context: densification in networks follows power law growth patterns due to the fact that the number of nodes and edges grows as power laws.

This class of models also includes a recent variant called the NetTide model[67], which describes power law growth patterns in the number of users in social networking sites, such as WeChat and Weibo. The NetTide model focus on a single product, where existing users invite non-users with certain time-varying probability: $\frac{dI_i}{dt} = \frac{\beta_i}{t^\theta} I_i(t)(I_i^\infty - I_i(t))$.

By setting $\theta = 1$, the model predicts that the early growth pattern of a product follows a power law: $I \propto t^\beta$. Similar equations have also been proposed to understand technology penetration[68,69]. This class of network densification models usually focuses on the growth patterns of one single product. While it is not clear how, and if at all, one may generalize the model to describe systems containing multiple products, the ability of these models to predict power law growth raises an interesting question: how do they relate to the observed power law patterns documented in our paper? Next we show, the growth patterns predicted in network models described in this section pertain to fundamentally different processes than what we observed, hence can not be adapted to explain our phenomena.

26

(**Relationship Between the *MS* Model and Network Growth Models**) The key for the two classes of network models described above to generate power law growth is because of the growth of the system. That is, the number of nodes and edges increases with time as a power law.

This raises an interesting question: Can the power law growth pattern we observed in substitutive systems be explained by the expansion of the system? Indeed, as we show in Supplementary Figure 16, while our system converges quickly to a relatively stable system, it still grows slightly over time with addition of new users. To answer this question, we study a stable system by removing the contributions from new subscribers in our dataset.

The new system is comprised of 1.64 Million people and their usage patterns in a two-year time window from 2010 to 2012. Each individual uses only one product at a time in this period (Supplementary Figure 16). For each of the product released in the two year period, we define its impact $I(t)$ as the total number of users among the population. Because there is no growth in the number of users in our system, the network growth models described above would break down, which raises an interesting question: would the power law growth persist in the absence of system growth? After eliminating the effect of growth in the number of users, we find the impact dynamics of individual handsets remain intact, following again a clear power law growth pattern (Supplementary Figure 19). Two sample Kolmogorov-Smirnov (KS) test shows that the distribution of $\eta$ presented here is no significantly different from the distribution shown in Fig. 1M ($p = 0.2842$, much larger than 0.05). This finding indicates that the observed power law growth pattern is not due to the growth of the system. Rather it pertains to mechanisms that operate within

systems.

In Supplementary Note 4, we present in detail our Minimal Substitution (*MS*) model. The model not only allows us to offer a mechanistic explanation for the observed power law growth pattern in substitutive systems, but also accurately captures the entire lifecycle of product impacts, collapsing constituents from a wide range of domains into a single universal curve, documenting a remarkable degree of regularity underlying the ubiquitous substitutive systems.

**Collective Behavior Models**

Collective Behavior Models are another line of important works dealing with processes such as collective online behavior, collective attention competition and bursty dynamics in human society. Next, we will first focus on two seminal works by reviewing the early growth pattern predicted by them, followed by discussing the relationship between the *MS* model and the bursty dynamic models.

(**Product Competition Models**) In 2007, Wu and Huberman proposed a model to describe how attention to novel items propagates and fades among large populations[70]. In their modeling framework, the dynamic of the popularity of top items can be described as:

$$I(t) = \prod_{s=1}^{t}(1 + r_s X_s)I(0) \approx \prod_{s=1}^{t} e^{r_s X_s} I(0) = e^{\sum_{s=1}^{t} r_s X_s I(0)}, \qquad (26)$$

where $r_s$ captures the temporal decay factor, $I(0)$ captures the initial influence, and $X_s$ are positive independent random variables. Wu and Huberman found that $r_s$ decays as a stretched exponential function, which can be described as: $r_t \sim e^{-0.4t^{0.4}}$. By inserting it back to (26), we can see that the

28

early growth pattern for $I(t)$ following exponential growth pattern, indicating that the model is not sufficient to explain the observed power law growth patterns in substitutive systems.

In another important work[71], Glesson et al. compared two possible mechanisms in product decision process by exploring a dataset of Facebook apps, the cumulative rule (where users make decisions based on cumulative sales) and recent activity rule (where users make decisions based on recent adoptions of apps). They found that the activity rule combined with a long-memory function offers a better fit of the data, indicating that people focus more on recent activities. Mathematically, to quantify the recent activity function, they define:

$$p_i^r(t) = L \sum_{\tau=0}^{t-1} W(t,\tau) f_i(\tau), \qquad (27)$$

where $f_i(t) = I_i(t) - I_i(t-1)$, captures the increment of the app influence. $W(t,\tau)$ corresponds to a memory function, determining the weight of the activity. Here they selected the exponential decay function: $W(t,\tau) = (1/T)e^{-(t-\tau)/T}$. They find that only the recent-activity rule with larger $T$ ($T = 50$) can reproduce the macroscopic properties of the data, while the cumulative rule or recent-activity rule with smaller $T$ ($T = 5$) reproduces only a part of the observations. By incorporating (27) into a simple growth function, we find that power law growth pattern cannot be a solution to the system, indicating that the model is also not sufficient to explain the observed growth patterns.

(**Bursty Human Dynamic Models**) The inter-event time distribution of several human actions are following fat-tailed distributions[72–75]. Indeed, if the inter-event time distribution $P(\Delta t)$ follows a fat-tailed distribution, it by itself could lead to a non-exponential early growth pattern,

29

which has been observed in several literatures[37,76–78]. This raises an interesting question, whether the observed power law early growth pattern is generated by bursty behavior in product purchases? To test this hypothesis, we measure $P(\Delta t)$ in the handset dataset, where $\Delta t$ captures the inter-event time between two purchases of one user (Supplementary Figure 20). Interestingly, we find that the inter-event time distribution follows a narrow distribution, best approximated by an exponential tail $P(\Delta t) \sim e^{-0.0025\Delta t}$. We further use a canonical method for testing power law distributions (by Clauset et al. in a seminal paper in 2007[79]). Here we test three different distributions, 1) $P(\Delta t)$ with $\Delta t < 100$, 2) $P(\Delta t)$ with $\Delta t < 500$ and 3) $P(\Delta t)$ with all data points included, finding that for all these distributions, $p < 10^{-3}$, which rules out the possibility that the waiting time distribution may be described by power law distributions. Therefore, although bursty human dynamics could be a simple explanation for the observed temporal patterns, Supplementary Figure 20 shows directly that its underlying assumption is unfortunately invalid, forcing us to exploring other mechanisms as we did in this work.

## Supplementary Note 4: Minimal Substitution (*MS*) Model

**Model Description**

In the proposed model, we consider a conservative system comprised of $N_0$ users, where each individual uses one handset at a time. Note that, the average number of products per user does not have to be around one. With time, new handsets are introduced into the system at a constant rate

$\rho$, prompting users to substitute their incumbent products with new innovations. In each time step, an individual substitutes another handset $j$ for her/his current handset $i$ with probability $\Pi_{i \to j}$:

$$\Pi_{i \to j}(t) = \lambda_{ij} N_j(t) \frac{1}{t_j}, \qquad (28)$$

where $N_j(t)$ is a time-dependent factor, measuring the popularity of the handset $j$ at time $t$ and $t_j$ measures its current age. The factor $N_j(t)$ in (28) captures the *preferential attachment* mechanism[15–17], suggesting that people tend to adopt handsets of higher popularity. The $1/t_j$ factor corresponds to the *recency* mechanism, uncovered by the data collapse documented in Fig. 3E. Indeed, while two-years is the typical age of a handset when it is substituted by other products, the distribution of the age of substitutes peaks much earlier (Supplementary Figure 22), indicating that user prefers handsets that are released more recently. The factor $\lambda_{i \to j}$ reflects the inherent *propensity* between two given products $i$ and $j$, capturing the heterogeneous nature in the likelihood of substitutions. Note that all factors in (28) are empirically validated and motivated in the main text, hence (28) represents a minimal model that brings together all mechanisms we know to date governing substitutions.

Indeed, there are many exogenous variables that may affect substitution dynamics in the handset system. For instance, the price and features of various products may influence a user's decisions; The existence of subscription plans in the mobile phone settings, including the durations and pricing structures of such plans, may also affect substitution dynamics. How precisely these exogenous features are correlated with the fundamental parameters we derive with the *MS* model remains an open question. But as we show in this work, by just considering these three simple parameters, we are able to not only analytically predict the observed power law growth patterns in

31

the early stage, but also accurately captures the trajectories of individual items in the systems.

Another interesting question is whether our MS model can only capture cases where the new product is better than the incumbent. Equation (28) predicts that a user is most likely to switch from an elder model 1 to newer model 2 (i.e., $T_1 < T_2$ where $T_1$, $T_2$ are the releasing time for handset model 1 and 2, respectively) if the propensity parameters are comparable ($\lambda_{1\to 2} \approx \lambda_{2\to 1}$). Yet, the probabilistic nature of the model indicates that it also allows the possibility for reverse switching from handset 2 to 1, especially if the two handsets were not released too far apart ($T_1$ is close to $T_2$) and $\lambda_{2\to 1} > \lambda_{1\to 2}$, indicating that the *MS* model is flexible and can be easily extended to capture reverse flows from an newer product to an old one.

**Solving the *MS* Model**

Given (28), the popularity dynamics of an individual handset can be expressed in the master equation formalism:

$$\begin{aligned}\frac{dN_i}{dt_i} &= \sum_k \Pi_{k\to i} N_k - \sum_j \Pi_{i\to j} N_i \\ &= \sum_k \lambda_{k\to i} N_k N_i t_i^{-1} - \sum_j \lambda_{i\to j} N_i N_j t_j^{-1}.\end{aligned} \quad (29)$$

Defining fitness as $\eta_i \equiv \sum_k \lambda_{k\to i} N_k$ and longevity $\tau_i$ as $\tau_i \equiv 1/\sum_j \Pi_{i\to j}$ ( the time-independence of the parameters will be proved in the next section). we have

$$\frac{dN_i}{dt_i} = \eta_i N_i t_i^{-1} - N_i/\tau_i. \quad (30)$$

By change of variable $f_i = \ln N_i$, we rewrite (30) as:

$$\frac{df_i}{dt_i} = \eta_i t_i^{-1} - 1/\tau_i. \tag{31}$$

By solving the equation, we arrive at:

$$f_i = \eta_i \ln(t_i) - t_i/\tau_i + C_i, \tag{32}$$

and

$$N_i(t) = h_i t^{\eta_i} e^{-t/\tau_i}. \tag{33}$$

Here, $h_i \equiv e^{C_i}$ corresponds to the anticipation factor. Since the impact of a handset ($I$) measures its total number of adopters, the impact dynamics can be obtained by solving the following equation:

$$\frac{dI_i(t)}{dt} = \eta_i N_i(t) t^{-1}. \tag{34}$$

By inserting (33) into (34), we obtain:

$$I_i(t) = \int_0^t h_i \eta_i t^{\eta_i - 1} e^{-t/\tau_i} dt = h_i \eta_i \tau_i^{\eta_i} \gamma_{\eta_i}(t/\tau_i), \tag{35}$$

where $\gamma$ corresponds to the incomplete gamma function $\gamma_z(t) \equiv \int_0^t x^{z-1} e^{-x} dx$. Interestingly, (35) suggests that the impact of a handset should saturate to a constant. Indeed, If we take the limit $t \to \infty$, the formula predicts the ultimate impact of a handset:

$$I_i^\infty = h_i \Gamma(\eta_i + 1) \tau_i^{\eta_i}, \tag{36}$$

where $\Gamma(z) \equiv \int_0^\infty x^{z-1} e^{-x} dx$ is the gamma function.

**Time-independence of the Parameters**

In this section, we show that both $\eta_i$ and $\tau_i$ are time-independent parameters in a stationary system (Fig. 3A). Because the propensity parameter $\lambda_{k \to i}$ between two products is independent of the popularity of a product, i.e. $\lambda_{k \to i}$ is independent of $N_k$ and $N_i$, allowing us to write:

$$\begin{aligned}
\eta_i &\equiv \sum_k \lambda_{k \to i} N_k \\
&\approx \sum_k \lambda_{k \to i} p(\lambda_{k \to i}|i) \sum_k N_k \\
&= N_0 \sum_k \lambda_{k \to i} p(\lambda_{k \to i}|i) = N_0 \lambda_i^{\leftarrow}.
\end{aligned} \quad (37)$$

We discover that $\eta_i$ only depends on two time-independent parameters: $N_0$, the total number of users in the system, and $\lambda_i^{\leftarrow}$, the average propensity from all other handsets *towards* $i$, indicating that $\eta_i$ is also time-independent.

We repeat the calculations above for the longevity $\tau$, obtaining:

$$\begin{aligned}
1/\tau_i &\equiv \sum_j \Pi_{i \to j} \\
&= \sum_j \lambda_{i \to j} N_j t_j^{-1} \\
&\approx \sum_j N_j t_j^{-1} \sum_j \lambda_{i \to j} p(\lambda_{i \to j}|i) \\
&= \lambda_i^{\to} \sum_j N_j t_j^{-1}.
\end{aligned} \quad (38)$$

Note that $\sum_j N_j t_j^{-1}$ converges to a constant in a stationary system, allowing us to define $M_0 \equiv$

$\sum_j N_j t_j^{-1}$. we obtain:

$$1/\tau_i \equiv \sum_j \Pi_{i \to j}$$
$$\approx \sum_j N_j t_j^{-1} \sum_j \lambda_{i \to j} p(\lambda_{i \to j}|i) \qquad (39)$$
$$= M_0 \lambda_i^{\to}.$$

Eq. (39) reveals that the longivety $\tau_i$ is inversely proportional to two time-independent parameters: $M_0$, a global parameter capturing the *effective popularity* of handsets in the system and $\lambda_i^{\to}$, the average propensity *from* $i$ to all other handsets, thus demonstrating the time-independency of $\tau_i$. Note that we have made approximations in (37) and (39), by assuming that $\lambda$ and $N$ are independent. Next, we will demonstrate the time-independence of the parameters without making these approximations by studying a continuous formalism of the *MS* model.

Note that, to derive the master equation, the average number of products per user does not have to be around one as we have observed for handsets (Supplementary Figure 16). For convenience, let us call the average number of products per user as "cardinality". From our model, it can be shown that as long as cardinality is small, the substitutive dynamics we studied here remain the same. For example, if average household has two cars (cardinality= 2), we can simply treat each household as two separated individuals in the system, when tracing the substitution pattern for each item.

**Mapping the System into a Continuous Space**

In this section, we discuss a continuous formalism of the *MS* model by mapping handsets

35

into a property space, enabling us to rigorously show the time-independent nature of the model parameters. To do this, we introduce a continuous vector $\varphi$ to represent a given handset's functions and properties. For any handset in the system, we assume its growth dynamic is determined by $\varphi$. Hence the product's popularity could be denoted by $N(\varphi, t)$, with $t$ capturing the handset's current age, and $\varphi$ corresponding to its properties. In this continuous framework, an individual substitutes a handset $(\varphi', t')$ for another product $(\varphi, t)$ with probability:

$$\Pi(\varphi, t \to \varphi', t') = \lambda(\varphi, \varphi') N(\varphi', t') t'^{-1}, \tag{40}$$

where $\lambda$ is a function of $\varphi$ and $\varphi'$, capturing the propensity between the products. Since the total number of people in the system is a constant ($N_0$), the popularity of the handsets in the system satisfies the following condition:

$$N_0 = \rho \int_\varphi p(\varphi) d\varphi \int_0^\infty N(\varphi, t) dt, \tag{41}$$

where $\rho$ measures the release rate of new handsets and $p(\varphi)$ corresponds to a distribution, from which a new handset's $\varphi$ is drawn. The popularity dynamic of any individual handset follows the master equation:

$$\frac{\partial N(\varphi, t)}{\partial t} = \rho \int_{\varphi'} p(\varphi') d\varphi' \int_0^\infty [\Pi(\varphi', t' \to \varphi, t) N(\varphi', t') - \Pi(\varphi, t \to \varphi', t') N(\varphi, t)] dt. \tag{42}$$

Inserting (40) into (42), we have

$$\begin{aligned}\frac{\partial N(\varphi, t)}{\partial t} &= \rho N(\varphi, t) t^{-1} \int_{\varphi'} p(\varphi') d\varphi' \lambda(\varphi', \varphi) \int_0^\infty N(\varphi', t') dt' \\ &\quad - \rho N(\varphi, t) \int_{\varphi'} p(\varphi') d\varphi' \lambda(\varphi, \varphi') \int_0^\infty t'^{-1} N(\varphi', t') dt' \\ &= \eta(\varphi) N(\varphi, t) t^{-1} - N(\varphi, t)/\tau(\varphi), \end{aligned} \tag{43}$$

36

where the handset's fitness is defined as:

$$\eta(\varphi) \equiv \rho \int_{\varphi'} p(\varphi')d\varphi' \lambda(\varphi', \varphi) \int_0^\infty N(\varphi', t')dt', \qquad (44)$$

and its longevity as :

$$\tau(\varphi) \equiv \frac{1}{\rho \int_{\varphi'} p(\varphi')d\varphi' \lambda(\varphi, \varphi') \int_0^\infty t'^{-1} N(\varphi', t')dt'}. \qquad (45)$$

We find both parameters are time-independent and are only determined by $\varphi$.

To further test the time-independency of the parameters, we run an agent-based simulation of the minimal substitution model. To compare with real data, we reconstruct a system resembling the mobile phone dataset in terms of its time-scale and system size (Supplementary Figure 21). Specifically, we consider a conservative system comprised of 2.5M individuals, where new products are introduced with a constant rate. In each time step, each user substitutes another product $j$ for her/his current product of $i$ with probability $\Pi_{i \to j}$. The propensity parameter $\lambda_{i \to j}$ is drawn from a fixed distribution. We also set a small simulation time step (0.1) to investigate the relaxation period of the parameters, especially in the early region. To investigate the dynamical system generated by our agent-based simulation, we measure the early growth patterns of each individual products, finding that it adequately reconstructed the observed power law early growth patterns (Supplementary Figure 21A). To quantify how fast the quantities $\eta_i$ and $\tau_i$ reach stationary state after new products are introduced, we measure $\eta_i(t_i) = \sum_k \lambda_{k \to i} N_k$ and $\tau_i(t_i) = \sum_k \lambda_{k \to i} N_k$ as functions of the age of product $i$ ($t_i$) (Supplementary Figure 21BC). We find both quantities are rather stable over time, and reach stationarity relatively quickly - faster than the time scale we study. We further measure the distribution of the parameters for the same products at different age

37

($t_i = 0, 10, 90$) and use a two-sample KS test to analyze the curves, finding the distributions collapse to each other ($p > 0.1$), again demonstrating that the parameter estimations are not affected by stationarity (Supplementary Figure 21DE).

**Maximum Likelihood Estimation of Model Parameters**

In order to test the model performance, we need to estimate the best parameter set ($h$, $\eta$, $\tau$) of each product, simulate its impact dynamics through (35), and compare it with the empirical observation. To achieve this, let us imagine a non-homogeneous stochastic process $\{x(t)\}$, with $x(t)$ representing the number of new adoptions by time $t$, satisfying:

$$Prob(x(t+h) - x(t) = 1) = \lambda_0(x,t)h + O(h^2), \tag{46}$$

where $\lambda_0(x,t)$ is a time dependent rate parameter. Given an empirically observed set of $N$ events $\{t_i\}$ within the time period $[0, T]$, where $t_i$ indicates the moment when the product gets adopted the $i^{th}$ time, the likelihood that the product's impact dynamics follows can be evaluated by the log-likelihood function:

$$\begin{aligned}\ln L &= \sum_{i=1}^{N} \ln(\lambda_0(i-1, t_i)) - \int_0^T \lambda_0(x(t), t) dt \\ &= \sum_{i=1}^{N} \ln(\lambda_0(i-1, t_i)) - \sum_{i=0}^{N} \int_{t_i}^{t_{i+1}} \lambda_0(i, t) dt.\end{aligned} \tag{47}$$

To find $\lambda_0(x, t)$ in our system, we insert (33) into (34), yielding

$$\frac{dI_i}{dt} = h_i \eta_i t^{\eta_i - 1} e^{-t/\tau_i}. \tag{48}$$

Thus, in our system, we have $\lambda_0 = h\eta t^{\eta-1} e^{-t/\tau}$. By change of variable $H \equiv h\eta$ and $\nu \equiv 1/\tau$, we

38

obtain the log-likelihood function:

$$\ln L = N \ln H + \sum_{i=1}^{N}(\eta - 1)\ln(t_i) + \sum_{i=1}^{N}(-\nu t_i) - H\nu^{-\eta}\gamma_\eta(\nu T). \quad (49)$$

The best-fitted parameters should maximize the log-likelihood function, satisfying the following equations,

$$\frac{\partial \ln L}{\partial H} = 0$$
$$\frac{\partial \ln L}{\partial \eta} = 0 \quad (50)$$
$$\frac{\partial \ln L}{\partial \nu} = 0.$$

These equations lead to a set of non-linear equations,

$$H - N\nu^\eta \gamma_\eta^{-1}(\nu T) = 0$$
$$\sum_{i=1}^{N} \ln t_i + N \ln(\nu) - N\gamma_\eta^{-1}(\nu T) j_\eta(\nu T) = 0 \quad (51)$$
$$-\sum_{i=1}^{N} t_i + N\eta\nu^{-1} - N\gamma_\eta^{-1}(\nu T)T[(\nu T)^{\eta-1}e^{-\nu T}] = 0,$$

where $j_z(x) \equiv \partial \gamma_z(x)/\partial z$ is the partial derivative of the incomplete gamma function on $z$. By solving (51), we are able to obtain the best fitted set of parameters ($h$, $\eta$, $\tau$) for each product. Since the parameters are estimated jointly in our model, they are compatible with any correlations real systems might possess. While initial studies have shown interesting correlation between parameters, the correlations do not affect the conclusion presented in the paper (power law early growth and entire growth pattern), as they pertain to a higher-order characterization of the systems.

**Model Performance and Limitations**

We randomly select six handsets as examples to illustrate the model validation process. We learn the best fitted parameters $(h, \eta, \tau)$ for each of the product, insert them back into (35) and simulate the impact dynamic. We find the model not only well captures the early power growth pattern of each handset (Supplementary Figure 23A), but also accounts for their entire impact dynamics (Supplementary Figure 23B). In Supplementary Figure 23C, we show the impact trajectories of 100 different handsets, finding excellent agreement between the model predictions and empirical observations. The performance of the model does not rely on the particulars of the system. In Supplementary Figure 24A-C, we show the impact dynamics of 70 automobiles, 200 apps and 500 scientific fields. Again, the model captures impact trajectories accurately in both systems. To systematically study the performance of the model, we calculate the coefficient of determination ($R^2$) for each fitting in all four systems and show the complementary cumulative distributions in Supplementary Figure 24D, finding that the model accurately captures the impact trajectories for a vast majority of the products.

Note that the lower incomplete gamma function in (35) has the following property $\gamma_{\eta+1}(x) = \eta\gamma_\eta(x) - x^\eta e^{-x}$, allowing us to define a normalized impact $Q(t) \equiv (I(t)/h - \tau^\eta \gamma_{\eta+1}(t/\tau))e^{t/\tau}$. Inserting it back to (35), we expect $Q(t) = t^\eta$. In the main text, we have shown the relationship between the normalized impact $Q$ as a function of the normalized time $t^\eta$ for all fitted products in the four systems, finding that the curves mostly collapse onto the same curve (Fig. 4). Here, we focus on products with $R^2 > 0.9$ (Supplementary Figure 25A-D), finding that the number of the remained products is still considerable across four systems, where we find 546 handsets, 86 automobiles, 1370 apps and 4450 scientific fields, again corroborating our modeling framework.

40

Systematic fitting evaluation also supports this conclusion (Supplementary Figure 25E). Hence given the obvious diversity in the dynamical patterns across different products, we find the amount of regularity uncovered by the simple model to be quite interesting.

Although the model provides a rather good fit for a vast majority of the products, there are occasional cases where the model prediction deviates from the data (1.23% of handsets, 1.1% of automobiles, 2.1% of apps and 0.87% of scientific fields). In Supplementary Figure 23D, we show an example of such a case in the handset dataset, indicating that impact dynamics with sudden discontinuities can not be captured by our model. The discontinuities could be caused by several factors, from hardware and software upgrades to marketing efforts made by the company. Indeed, handset retailers may promote and run campaigns on various handset brands just like any other product they sell. They can change the price of a product when a new version is released. While understanding how such information may affect the dynamics could enhance our capability in describing the trajectories of such occasional products, unfortunately, we do not have access to such information in our decade-long dataset. But we also wish to note that, despite the model fails in capturing the trajectories of the occasional cases ($\sim 1\%$ -2% of the products), we find that most trajectories can be quite accurately described by the three parameters our simple model predicted, which implies that main external factors that may drive impact can be absorbed into the three parameters.

**Comparison with Canonical Models**

To compare our model with existing models, we selected a few canonical models, fitting

41

them to our data and comparing them directly to the performance of the proposed MS model.

First, we show visually the fit between various models and data, highlighting the conceptual difference these models offer. Specifically, we show the fits of our *MS* model and the fits of other traditional models including Logistic, Bass and Gompertz model. Supplementary Figure 26 demonstrates how other models, being analytical models, fail to predict the power law growth with varying non-integer exponents. Both the Logistic and Gompertz model predicts an exponential growth. The Bass model belongs to the class of models whose early growth can be approximated as linear function (also, the first term of Taylor expansion of an exponential function), but the dynamical exponents are strictly one and cannot be varied to account for non-integers. The main reason for the clear deviations of these models is that they are not designed to capture the substitutive processes we studied here. In contrast, our model fits well the entire growth trajectories.

Second, to compare directly the performance of our *MS* with other models, we computed the weighted KS test for the fits to quantify early deviations between the fit and data[1]:

$$D_i = max_{t \in [0,T]} \frac{|I_i^t - \tilde{I}_i^t|}{\sqrt{(1 + I_i^t)(I_i^T - I_i^t + 1)}}. \tag{52}$$

Here, a lower $D$ is expected for a better fit model. In Supplementary Figure 27, we show the distribution of the weighted KS measure for the four systems. We not only compare our model with the three traditional analytical models, but also with the NetTide model - a model which has been used in understanding technology penetration and network growth (See Supplementary Note 3), finding the MS model systematically outperforms all these models. Note that although NetTide model cannot outperform our *MS* model, it provides a better fit compared with other traditional

42

models. To better understand its performance, we provide a further comparison between it and the *MS* model (Supplementary Figure 28), finding a vast majority of products indeed prefer our *MS* model.

**Linking Short-term and Long-term Impacts**

The *MS* model offers an intriguing linkage between a product's short-term impact and its long-term impact. By taking the derivative of (33), we obtain the moment $t_i^*$ when the product's popularity reaches its peak,

$$t_i^* = \eta_i \tau_i. \tag{53}$$

By inserting (53) and (36) into (35), we discover that the handset's impact at $t^*$ (short-term impact) and its ultimate impact $I^\infty$ (long-term impact) can be connected by a simple equation:

$$\frac{I_i^\infty}{I_i(t_i^*)} = \Phi(\eta_i), \tag{54}$$

where $\Phi$ is a function of $\eta$, defined as $\Phi(\eta) \equiv \frac{\Gamma(\eta)}{\gamma_\eta(\eta)}$.

In order to test the formula empirically, we calculate the impact of each handset by 11/03/2014 (the last date in our dataset), denoting them as $I^l$. We specifically focus on 469 handsets whose $I^l$ are close enough to their estimated ultimate impacts, satisfying the criterion: $\frac{I^\infty - I^l}{I^\infty} \leq 5\kappa$, where we choose $\kappa = 0.02$. To correct for the difference between the ultimate impact and $I^l$, we rescale $I^l$ with $1 - \kappa$, obtaining an empirically estimated ultimate impact $I^e = I^l/(1 - \kappa)$. As for $I(t^*)$, we learn the three parameters $(h, \eta, \tau)$ for each product, calculate its $t^*$ through (53) and find its empirical impact at $t^*$.

In Supplementary Figure 29A, we show $I^e$ as a function of $I(t^*)$, finding they follow clear linear relationship, consistent with the prediction of (54). Furthermore, in Supplementary Figure 29B, we normalized $I^e$ by $I(t^*)$, showing the ratio as a function of $\eta$. We find the slight increase trend in $\Phi(\eta)$ as a function of $\eta$ is again accurately predicted by (54).

**Quantifying the Dynamics of the Substitution Flow**

Another key innovation of the Minimal Substitution model (*MS* model) is that it captures detailed dynamical information about the substitution flux $J_{i \to j}(t)$ between products, provided by $J_{i \to j} = \lambda_{i \to j} N_i(t) N_j(t) t^{-1}$. In contrast, most traditional models including Bass model, logistic model and recent NetTide models focus mainly on predicting the total amount of adoptions for each product and do not provide any information about pair-wise transition[18,41,67]. In fact, serving as the driving force in determining the rise-and-fall pattern of product popularities, the substitution flow dynamic is quite important in substitution systems. It is highly non-trivial to model the pairwise flux that is consistent the empirical data. In Supplementary Figure 30A, we take the substitution from SonyEricsson W595 to Apple iPhone 4S as an example, showing the dynamic of the substitution flow as a function of time. We fit our *SM* model with the empirical observation, finding that the model provides a rather accurate description of the substitution dynamics. We also compare our model with an existing model[2,80], defined as: $J'_{i \to j} = \lambda'_{i \to j} N_i(t) N_j(t)$, fitting it to the empirical data, finding that the existing model overestimates the substitution flow as time increases. To systematically compare the *MS* model to the existing model, we select three snapshots, plotting the fitted substitution flow ($\tilde{J}$) as a function of the real substitution flow ($J$) (Supplementary Fig-

ure 30B-D). We find the *MS* model provides a rather good fit of the substitution flow for all three different time snapshots (3 months, 6 months and 12 months), while the null model overestimates the substitution flow for these periods.

# Supplementary Figures

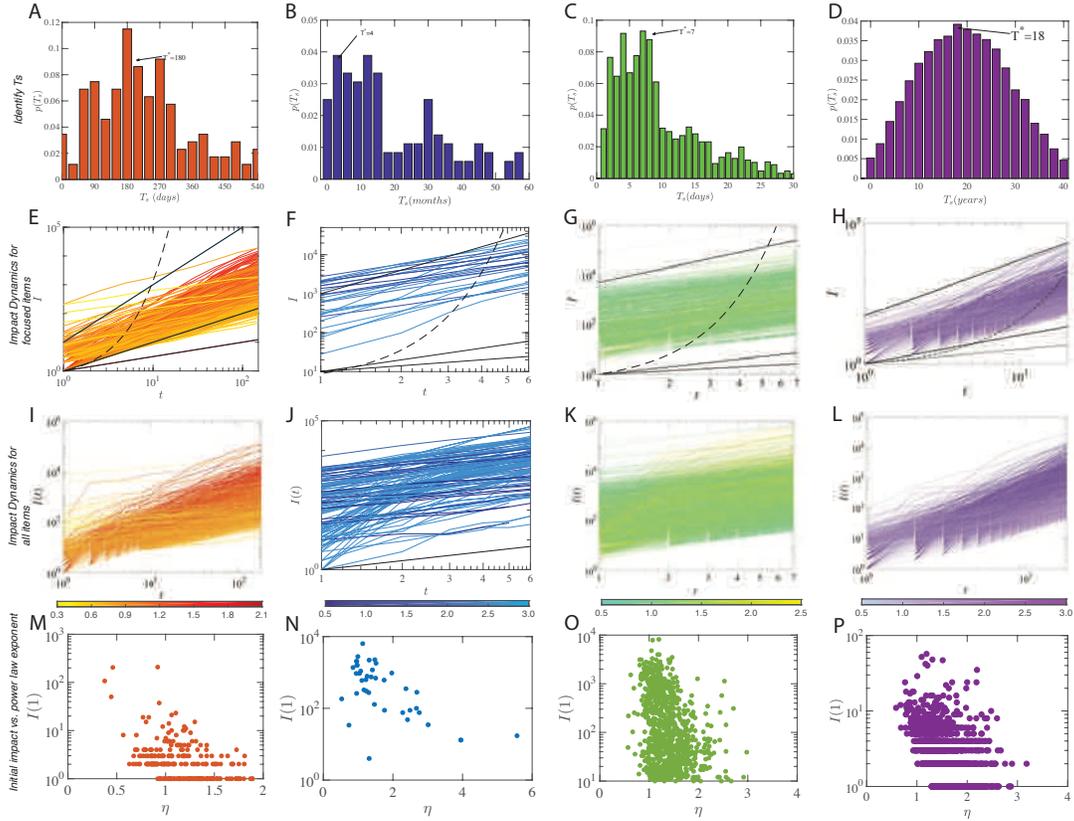

Supplementary Figure 1: **Dataset description. (A–D)** Distribution of $T_s$ for four datasets: handset (A), automobile (B), smartphone app (C) and scientific field (D). We identify $T^*$ as the position of the first highest peak of the distribution of $T_s$, finding $T^* = 180$ days for handsets, $T^* = 4$ months for automobiles, $T^* = 7$ days for apps and $T^* = 18$ years for scientific fields. **(E–H)** Impact as a function of time for focused items in datasets. The color of the line corresponds to the power law exponent of each handset. **(I–L)** Impact as a function of time for all items in four datasets. The color is coded by the slope of the power law hypothesis. **(M–P)** The impact of the first time unit (first day for handset and app dataset, first month for automobile dataset, first year for scientific field) as a function of the power law exponent $\eta$ characterizing the initial growth.

46

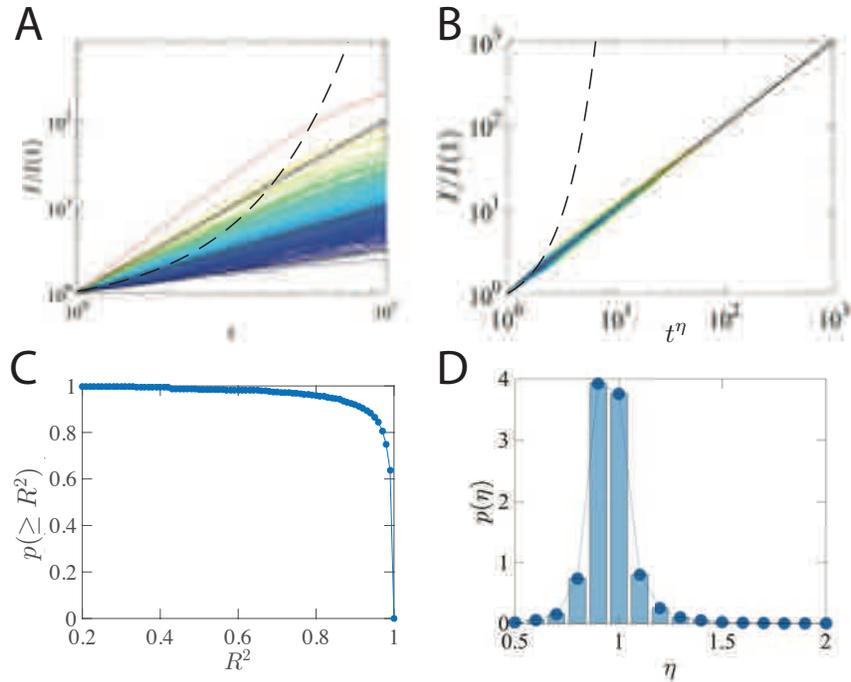

Supplementary Figure 2: **Health apps dataset.** (A) We repeat the analysis in Fig. 1 on the Health App dataset, finding the impact dynamics follow the same power law growth patterns: $I(t) \sim t^\eta$. The color of the line corresponds to the power law exponent of each handset. The solid black lines are $y = x^{1/2}$, $y = x$, and $y = x^2$, respectively; the dashed line corresponds to exponential growth, as guides to the eye. (B) We rescale the impact dynamics plotted in (A) by $t^\eta$, finding all curves collapse into y = x. (C) The complementary cumulative distribution of $R^2$, capturing how well the early growth patterns can be fitted as power laws. (D) Distribution of power law exponents $P(\eta)$ for curves in (A).

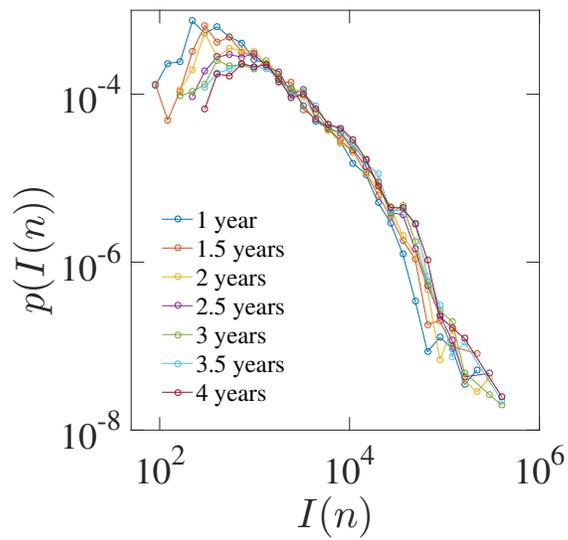

Supplementary Figure 3: **Distribution of Impacts**. Distribution of the $n$-year impact ($I(n)$) of handsets. $I(n)$ is defined as the number of sales of a handset after released for $n$ years.

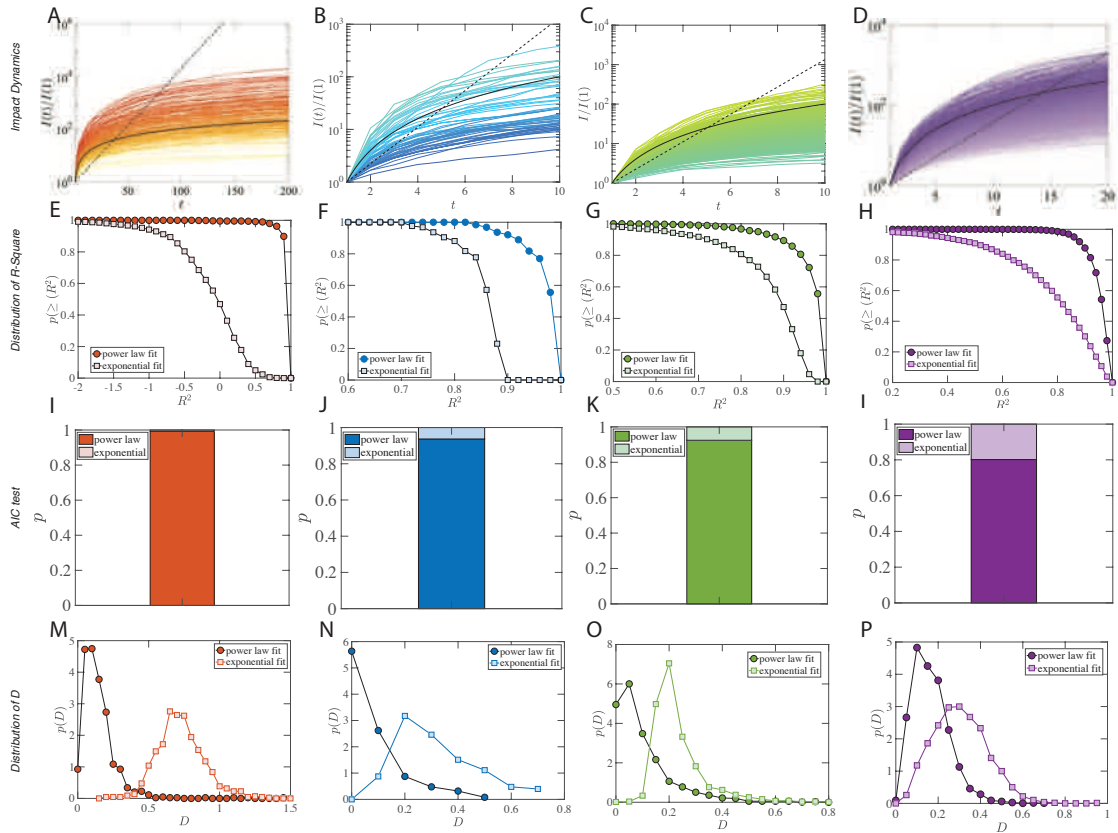

Supplementary Figure 4: **Power law versus Exponential fit**. **(A–D)** Normalized impact growth patterns in a semi-log plot for **(A)** Handset, **(B)** Automobiles, **(C)** Smartphone Apps and **(D)** Scientific Fields. Here the solid black curve corresponds to power law growth pattern and dashed line relates to exponential growth as guides to eyes. The products selected and the color code remain the same to Fig. 1 in the main text. **(E–H)** R-square test for the power law fit and exponential fit of entire sample. **(I–L)** Fraction of products which favor power law fit (exponential fit). **(M–P)** Weighted Kolmogorov-Smirnov (KS) test for the power law fit and exponential fit.

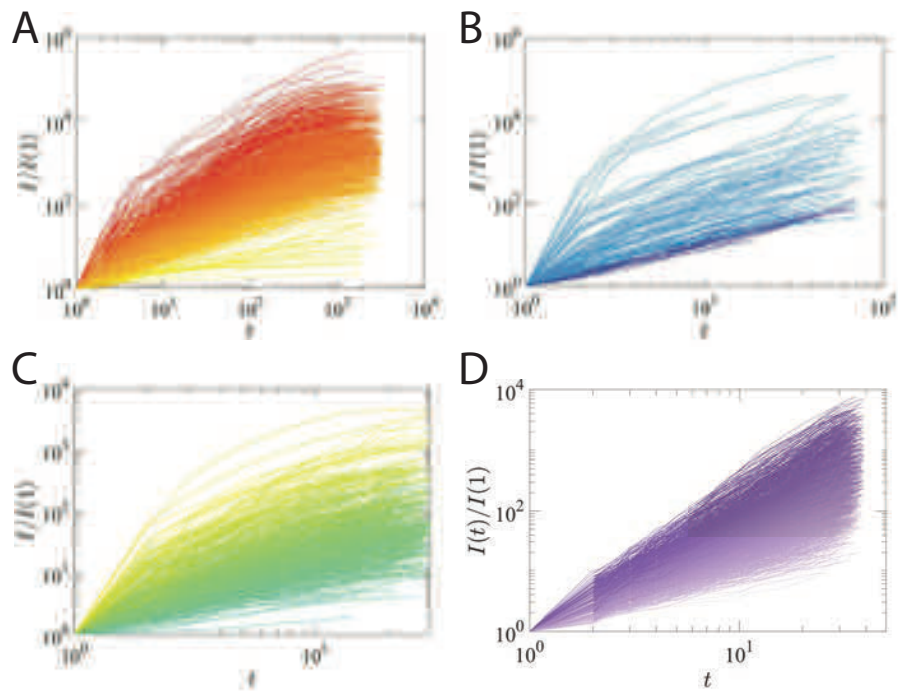

Supplementary Figure 5: **Impact growth patterns.** **(A–D)** The impact growth patterns for the entire lifecycle across the four datasets.

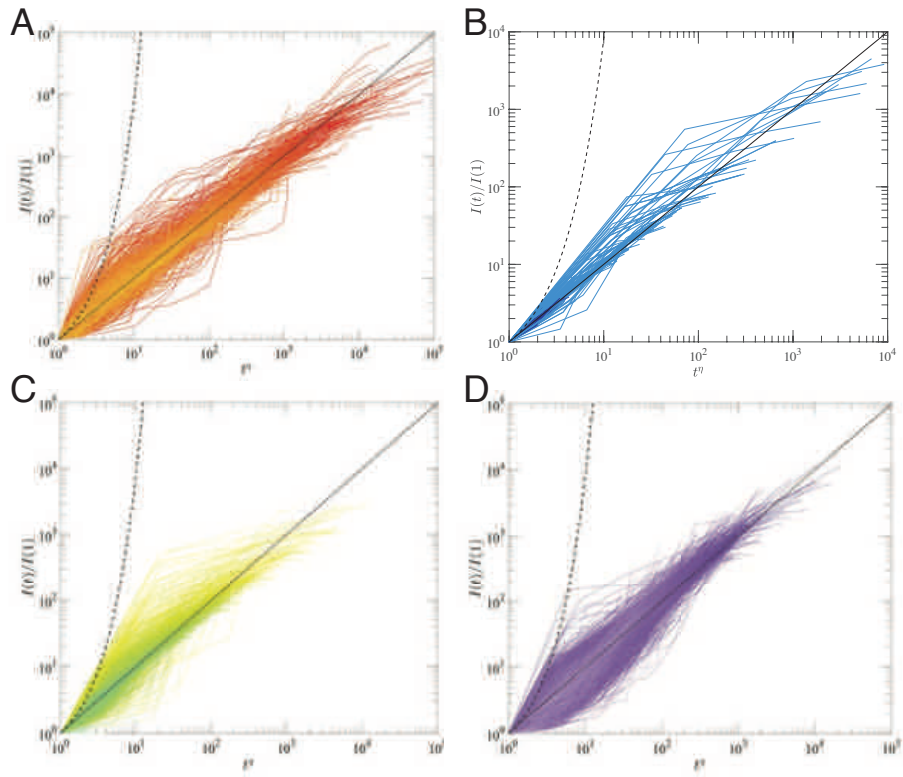

Supplementary Figure 6: **Rescaled impact growth patterns.** (A–D) The rescaled impact growth patterns for the entire lifecycle across the four datasets.

51

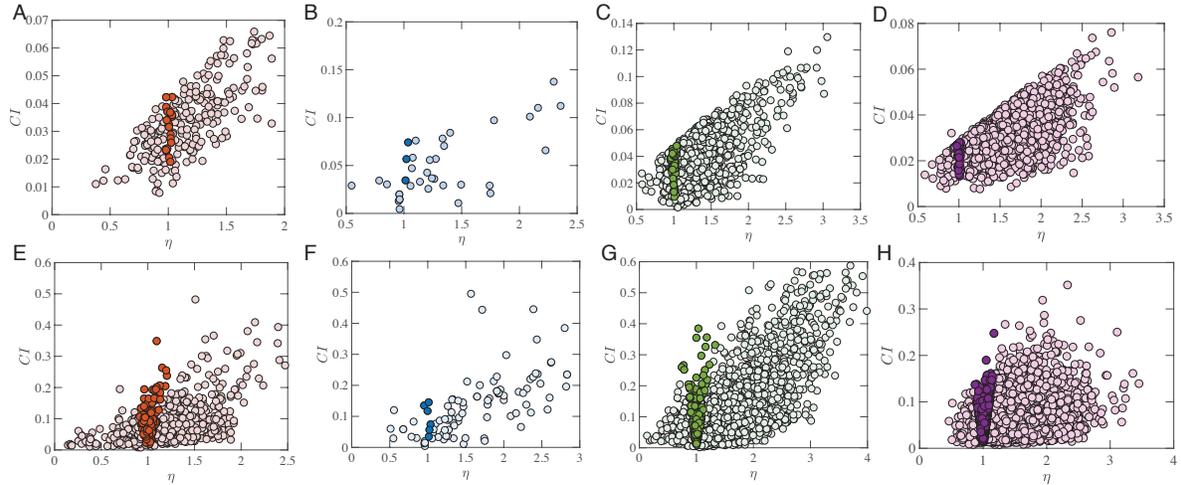

Supplementary Figure 7: **Confidence Intervals of the fittings**. **(A–D)** 95% Confidence Interval as a function of $\eta$ for products with products shown in Fig. 1. We find only a small fraction of products following linear growth pattern: 19 out of 240 handsets (7.9%), 3 out of 37 automobiles (8.1%), 59 out of 1,022 apps (5.77%), 105 out of 1,743 scientific fields (6.02%). **(E–H)** 95% Confidence Interval as a function of $\eta$ for products whose records are longer than $T^*$. We still find only a small fraction of products following linear growth pattern: 165 out of 885 handsets (18.6 %), 6 out of 119 automobiles (5.04 %), 214 out of 2,672 apps (8.01 %) and 396 out of 6,399 scientific fields (6.19%). Here we colored products with a darker color if they could be explained by a linear model.

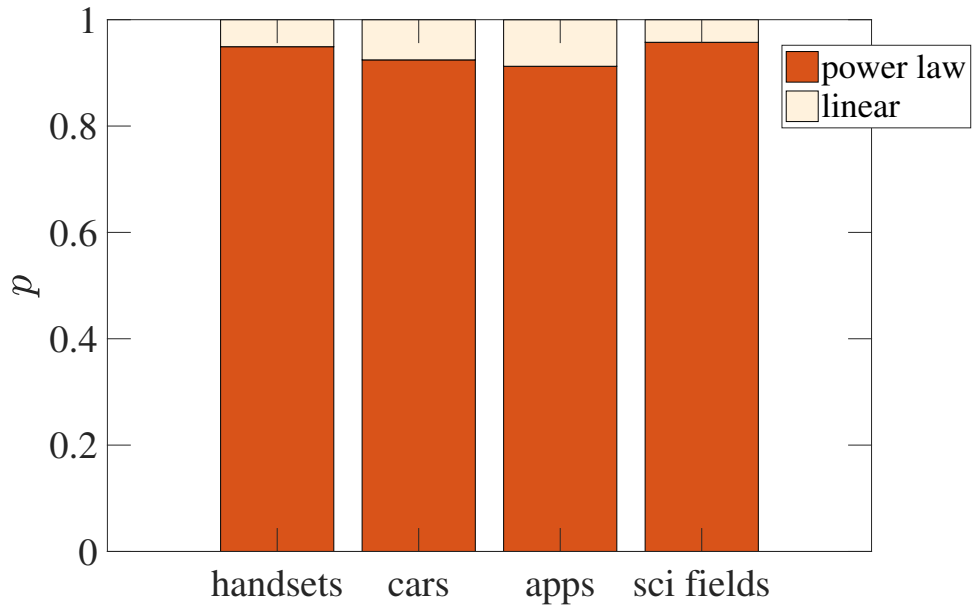

Supplementary Figure 8: **Power Law versus Linear fit**. Comparing power law growth with linear model for the early growth patterns of all handsets, automobiles, mobile apps and scientific fields with enough data points based on Akaike Information Criterion (AIC). We find, after excluding the influence from the additional parameter, the power law still performs better for a vast majority of the products, where only 45 out of 885 handsets (5.08%), 9 out of 119 automobiles (7.56%) and 234 out of 2,672 mobile apps (8.75%) and 272 out of 6,399 scientific fields (4.25 %) favor linear growth model.

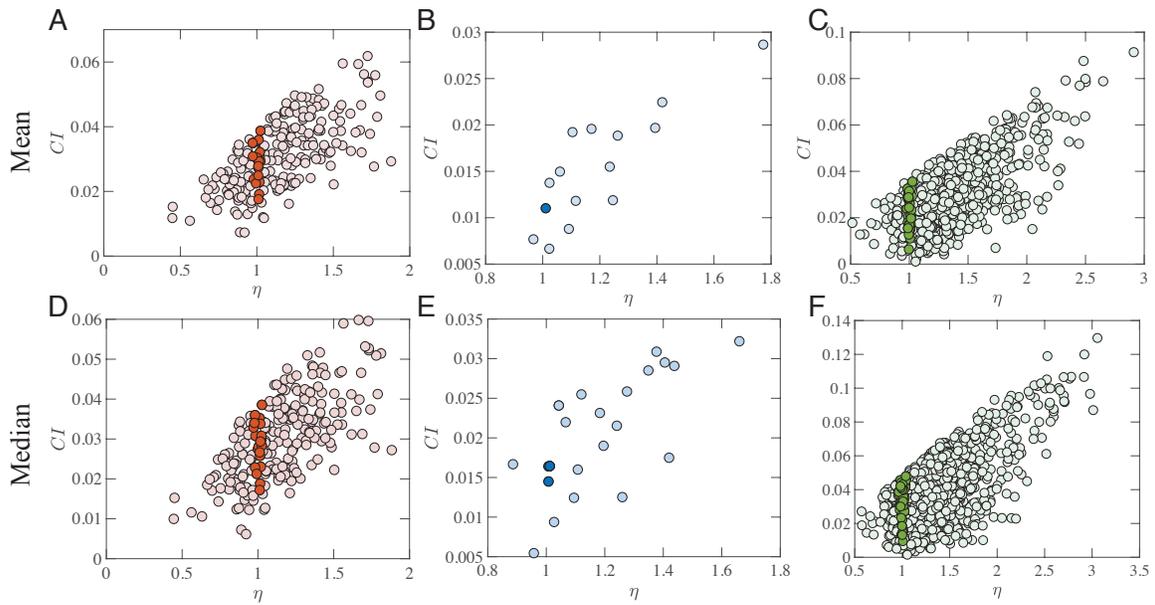

Supplementary Figure 9: **Confidence Intervals of the fittings for selected products (alternative definition for early growth phase)**. (A-C) 95% Confidence Interval as a function of $\eta$, for products with fitted $R^2 \geq 0.99$. Here we select the mean value of the distribution of $T_s$ to estimate the early growth phase. We find only a small fraction of products following linear growth pattern: 20 out of 237 handsets (8.44%, A), 1 out of 15 automobiles (6.67%, B), 32 out of 888 apps (3.6%, C). (D-F) Same measures to (A-C), but select median value to estimate early growth period. We find only a small fraction of products following linear growth pattern: 21 out of 240 handsets (8.75%, D), 2 out of 21 automobiles (9.95%, E), 59 out of 1,022 apps (5.77%, F). The scientific fields dataset is not shown here, since the median value and mean value of $T_s$ remain unchanged to the mode.

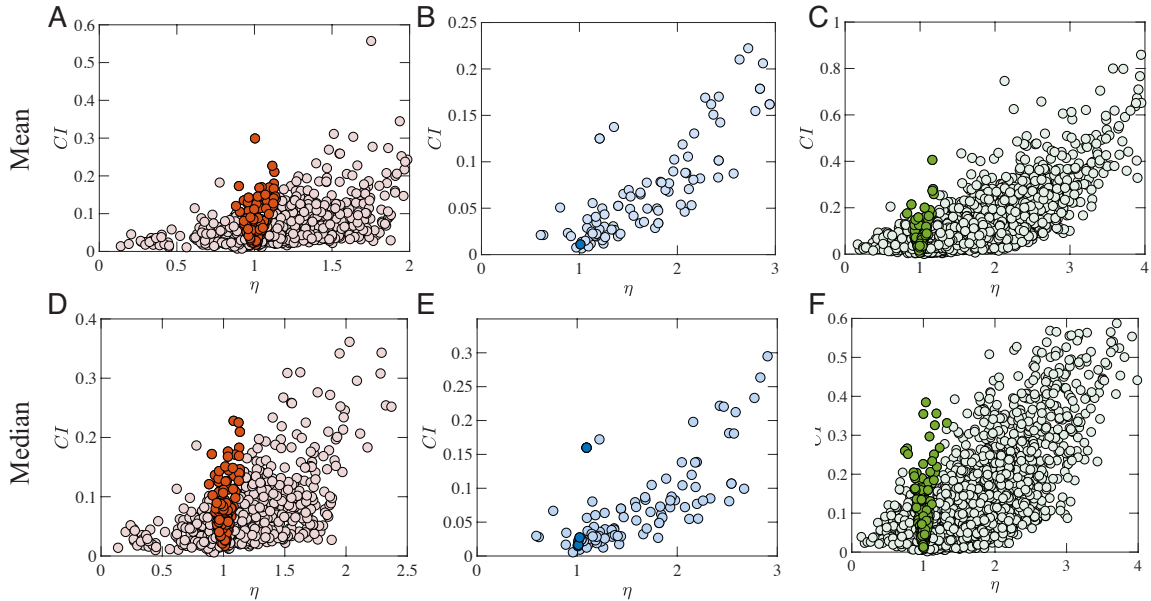

Supplementary Figure 10: **Confidence Intervals of the fittings (alternative definition for early growth phase)**. (A-C) 95% Confidence Interval as a function of $\eta$ for all products for whose records are longer than $T^*$. Here we select the mean value of the distribution of $T_s$ to estimate the early growth phase. We still find only a small fraction of products following linear growth pattern: 150 out of 856 handsets (17.52%), 1 out of 98 automobiles (1.02%), 170 out of 2,672 apps (6.36%). (D-F) Same measures to (A-C), but select median value to estimate early growth period. We find again 154 out of 869 handsets (17.72%), 8 out of 104 automobiles (7.69%), 214 out of 2,672 apps (8.01%). The scientific fields dataset is not shown here, since the median value and mean value of $T_s$ remain unchanged to the mode.

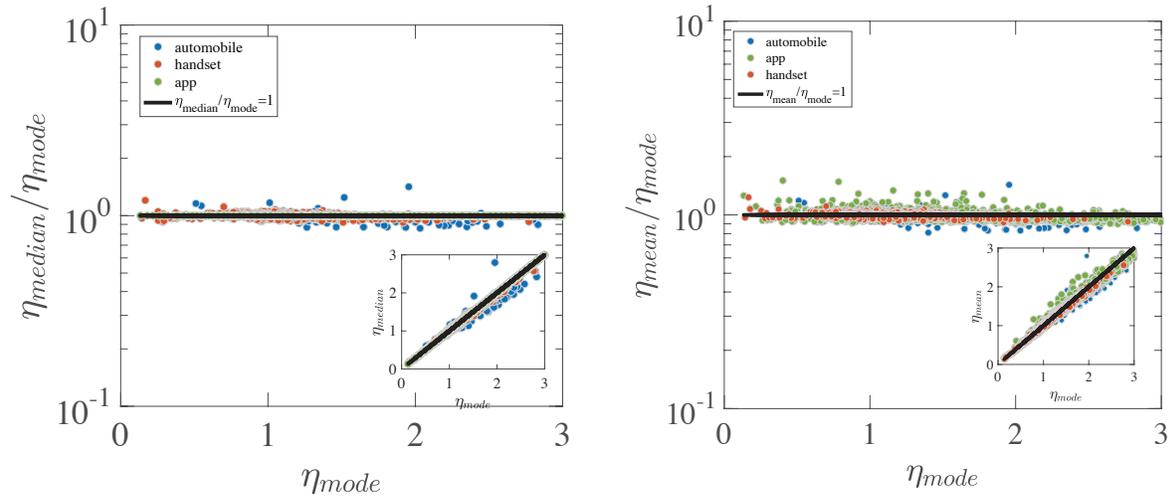

Supplementary Figure 11: **Robustness of the exponent fitting.** (A) The ratio between $\eta_{median}$ and $\eta_{mode}$ as a function of $\eta_{mode}$. (B) The ratio between $\eta_{mean}$ and $\eta_{mode}$ as a function of $\eta_{mode}$. We find for both definitions, the ratio remains to be a constant, indicating that the method is rather robust to definition selection. We also show $\eta_{median}$ and $\eta_{mean}$ as a function of $\eta_{mode}$ in the inset figures.

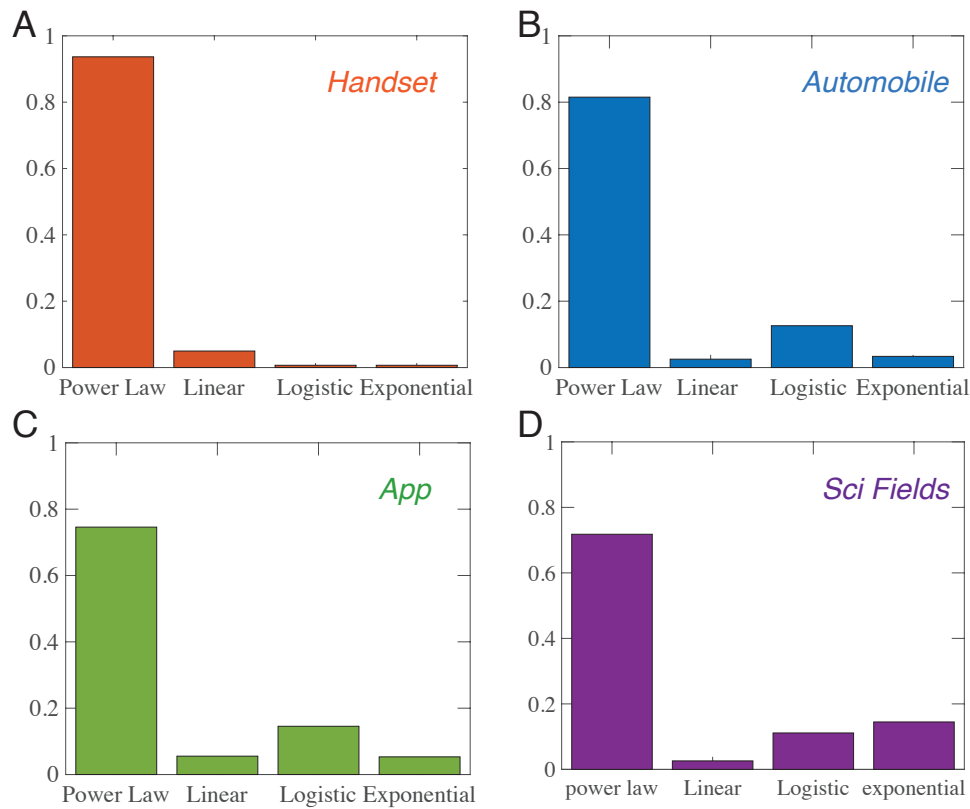

Supplementary Figure 12: **Comparison of Various Models.** (A-D) Comparing four various models (Power law, Linear, Logistic, Exponential) for all handsets (A), automobiles (B), mobile apps (C) and scientific fields (D) with enough data points. We find the growth pattern of 93.67% of handsets, 81.51% of automobiles, 74.59% of mobile apps and 71.79% of scientific fields (D) favor power law with non-integer exponents than other models. If we absorb the linear growth pattern into power law growth, we have 98.6% handsets, 83.5% automobiles, 79.6% apps and 74.1% scientific fields which favor power-law early growth patterns than exponential-class functions.

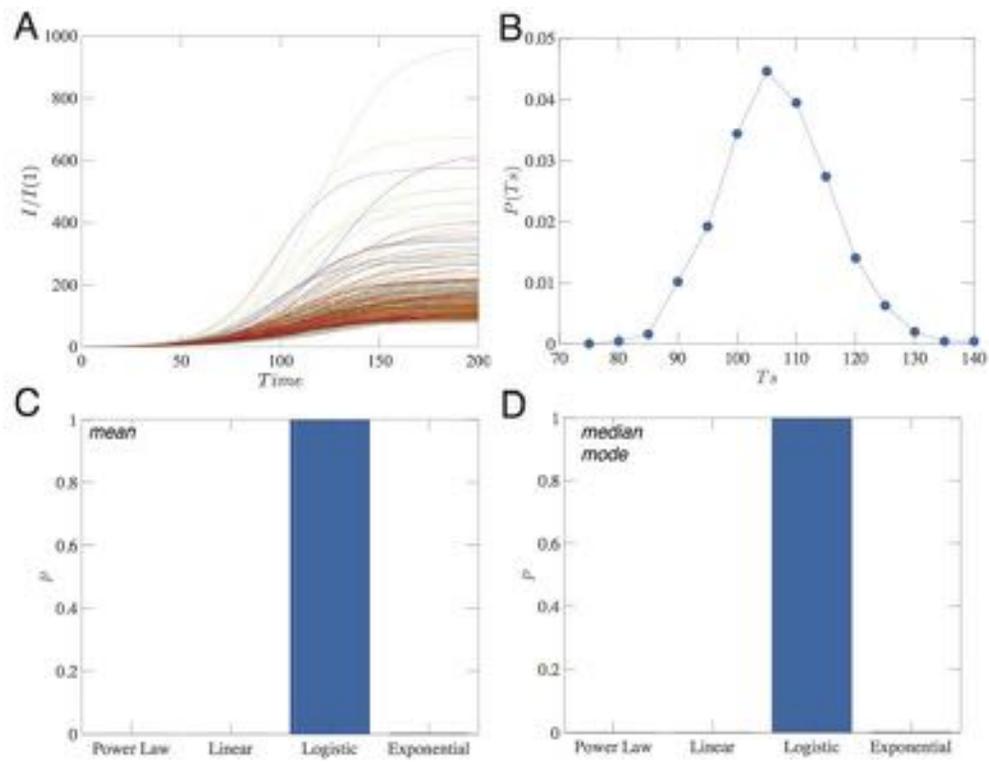

Supplementary Figure 13: **Testing robustness of the methods by using a logistic experiment**.

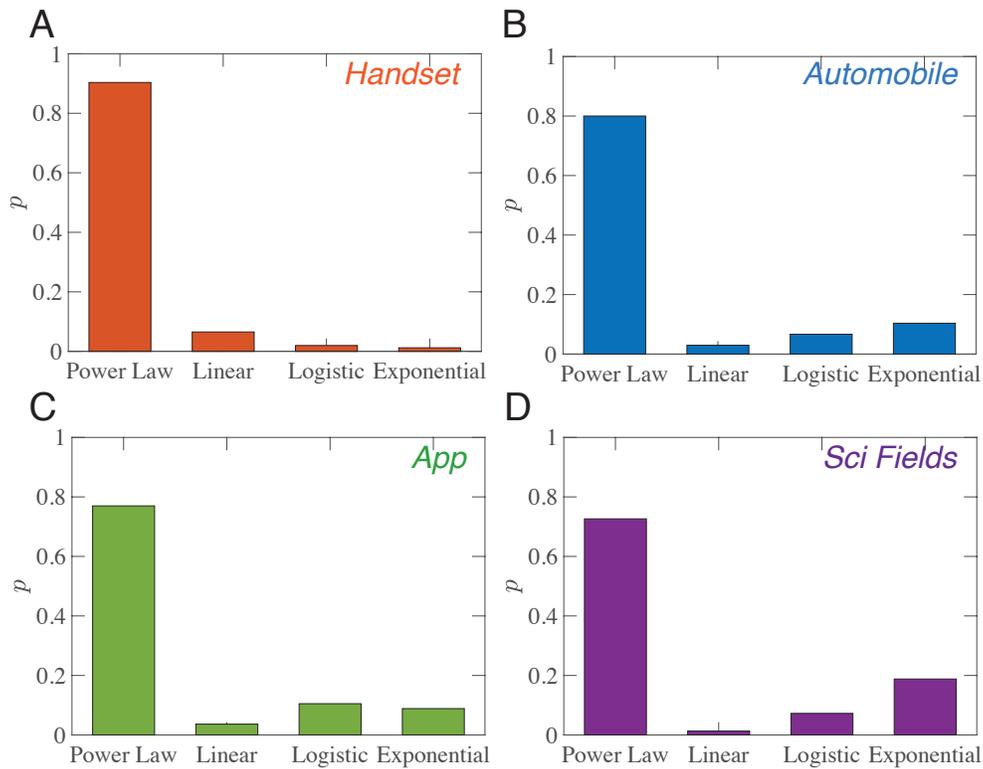

Supplementary Figure 14: **Comparison of Various Models (alternative definition of early growth phase)**. Testing the power law growth pattern with alternative definition of early growth period. Here we define $T^*$ as $T^* = T_s$ for each individual product, allowing them to have different early growth period. We find 90.35% of handsets (A), 80% of automobiles (B), 76.9% of mobile apps (C) and 72.63% of scientific fields (D) favor the power law models than other models.

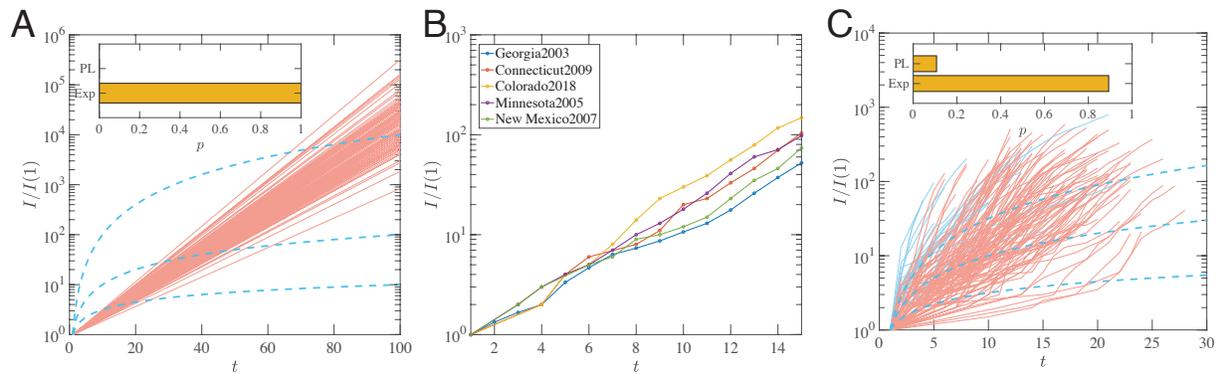

Supplementary Figure 15: **Method Validation on Flu Spreading Dataset (Non-substitution Dataset).** (A) 100 exponential curves with different exponents. The growth pattern follows a straight line in a semi-log plot, visually different from power law growth (dashed lines). Based on our method, 100% of curves have been classified as exponential instead of power law (inset). (B) The early growth pattern for 5 selected dynamics. We find for all of them prefer exponential growth. (C) The early growth pattern for all 168 cases of flu epidemics. We find 89.31% of the curves favor exponential than power law (inset). The curves that favor exponential growths are colored in red while the power law growth is colored in blue.

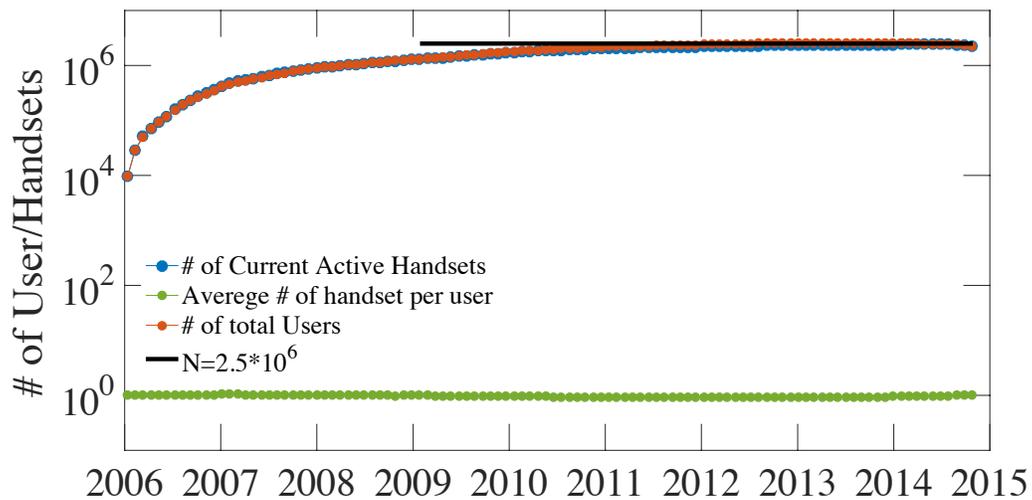

Supplementary Figure 16: **Handset system as a substitutive system**. Number of current active handsets (current active users) as a function of time. We find both quantities saturate to a constant (black line). We also calculate the average number of handsets per user as a function of time, finding that people are holding one single handset at a time.

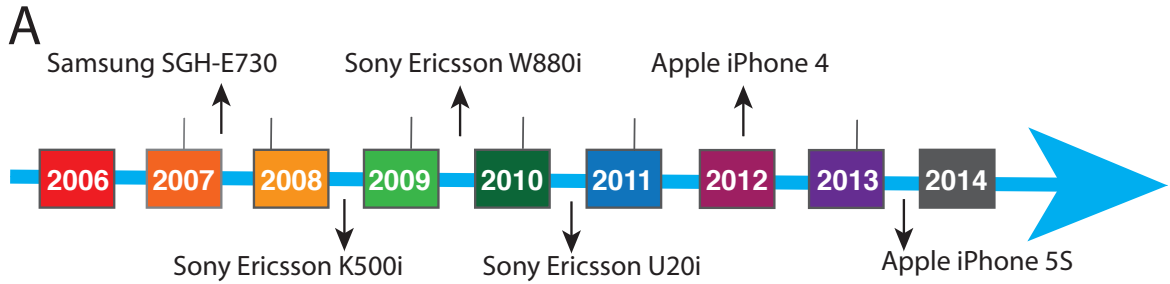

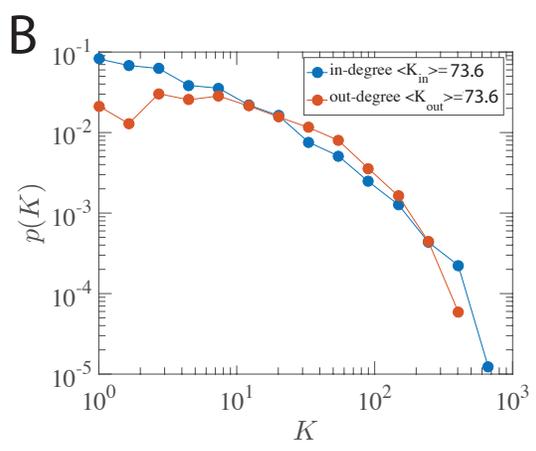
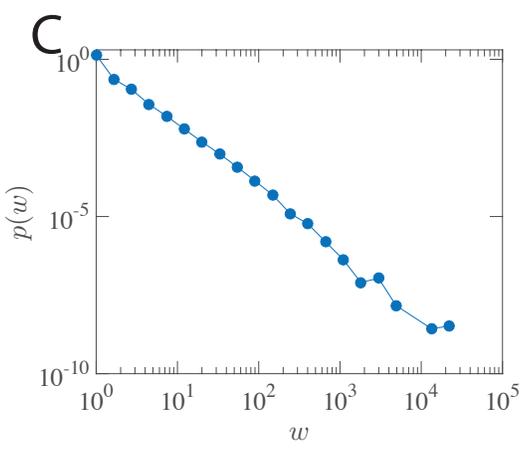

Supplementary Figure 17: **Substitution network**. (**A**) Illustration of a usage timeline. (**B**) In-degree and out-degree distribution of the aggregated substitution network generated between January 2014 and June 2014. (**C**) Link weight distribution of the substitution network.

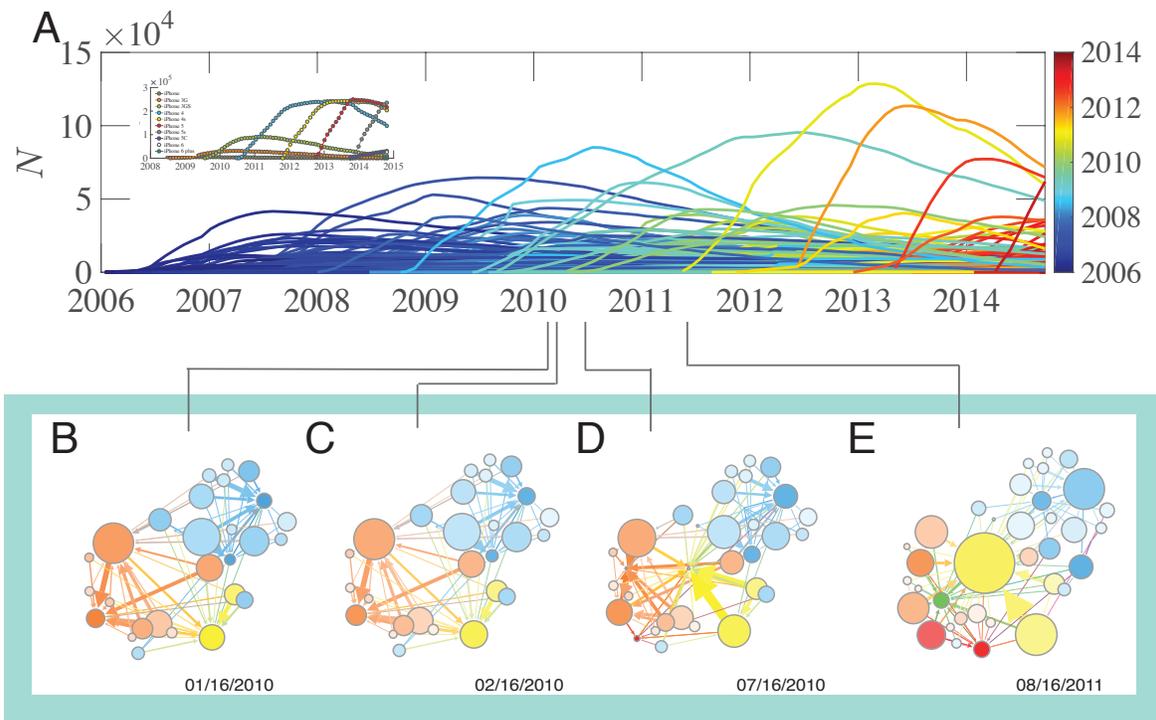

Supplementary Figure 18: **Characterizing substitution patterns**. (A) Popularity of individual handsets ($N$) over time for products from Apple Inc (inset) and products from other companies (main). Each line represents a model of handset. The color of the lines correspond to the release dates of the products, shifting from blue to red. (B–E) Substitution network of top handsets in four selected snapshots. We selected for handsets who were ranked within top 10 based on their popularity. The size of the nodes captures the popularity of the handset. Handsets by manufacturers are shown in different node colors, which fade with the age of handsets. The weight of the link captures the number of substitution in a period of one month. We find, in addition to the complexity and heterogeneity depicted in Fig. 2, substitution patterns are characterized by a remarkable amount of temporal variability.

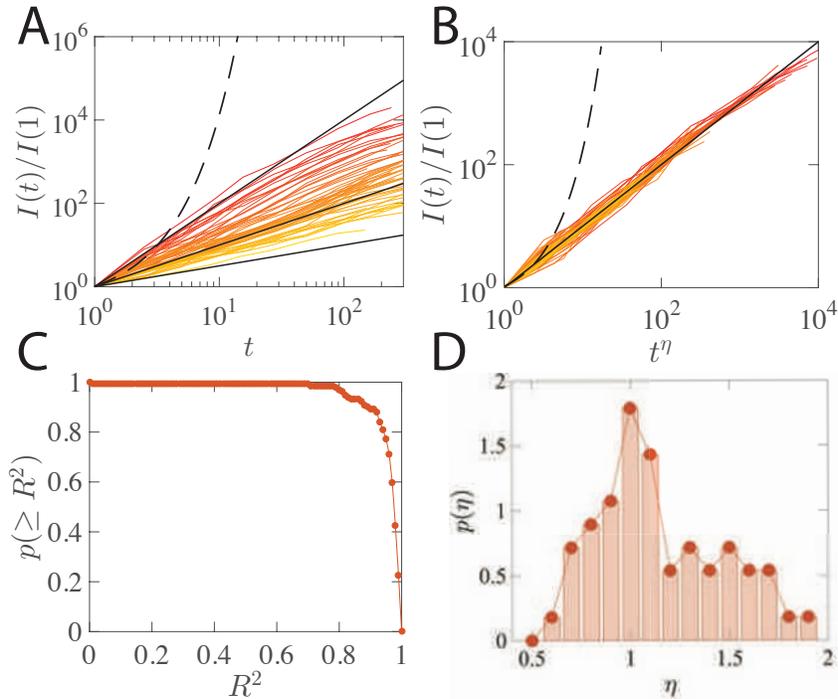

Supplementary Figure 19: **Power law growth persists when the number of users stays constant**. To eliminate the influence of the network growth on the power law growth pattern, we explore a conservative system comprised of 1.64 Million users in a two-year time window (2010-2012). By removing potential contributions from new users, existing network models (Supplementary Note 3) would predict the power law growth disappears. Yet, we find in our system the same power law growth patterns. **(A)** By repeating the analysis shown in Fig. 1, we study the growth pattern of 131 handsets released between 01/01/2010 and 06/01/2011 and selected 56 impact trajectories as power law examples. **(B)** We rescale the impact dynamics plotted in (A) by $t^\eta$, finding all curves collapse into y = x. **(C)** The complementary cumulative distribution of $R^2$, capturing how well the early growth patterns can be fitted as power laws. **(D)** Distribution of power law exponents $P(\eta)$ for curves in (A). Two-sample KS test shows that the distribution is no statistical different from Fig. 1M.

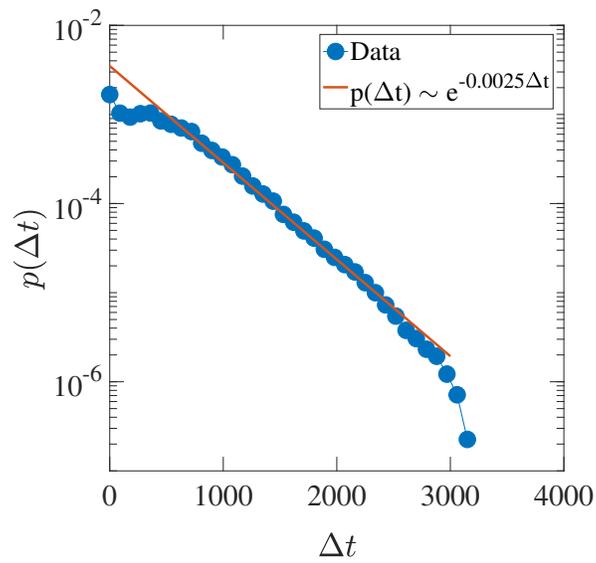

Supplementary Figure 20: **Inter-event time distribution in the handset dataset**. Here $\Delta t$ captures the time interval between two purchases for one user.

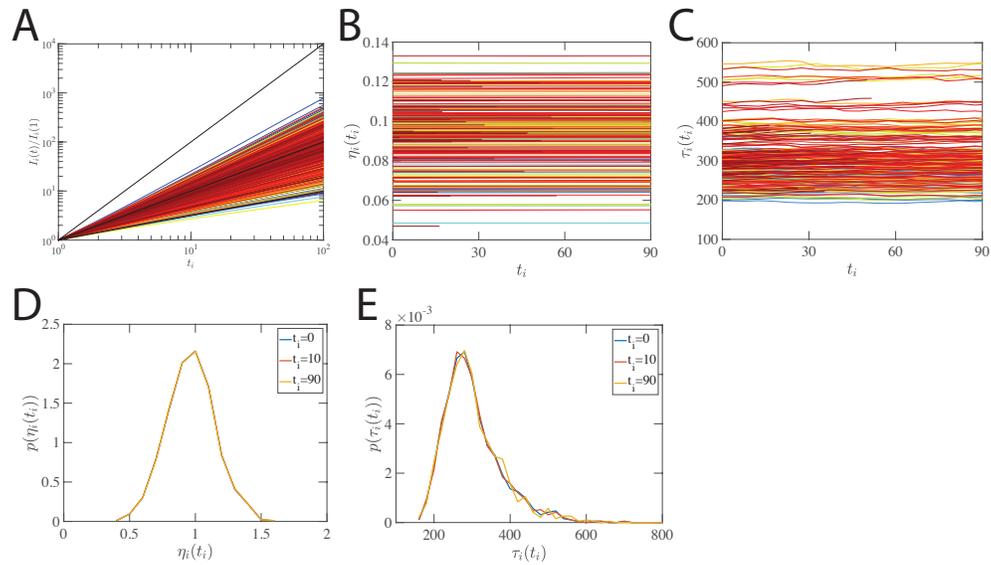

Supplementary Figure 21: **Time-independency of parameters.** (A) Early growth pattern of 1,000 products in the simulated system. (B-C) The dynamics of the parameters $\eta$ and $\tau$ as a function of time for 200 randomly selected products. (D-E) The distribution of $\eta$ and $\tau$ at different time age for all 1,000 products. We find the parameters are time-independent.

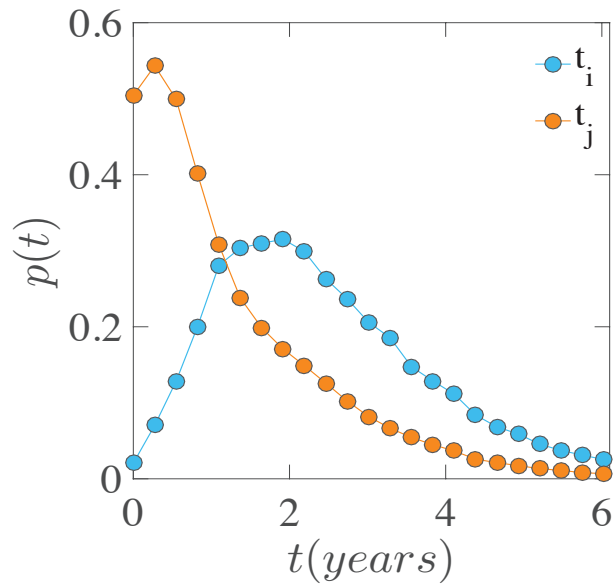

Supplementary Figure 22: **Distribution of handset age**. For each substitution event in the system, we measure the age of the substitutes ($t_j$) and the incumbent ($t_i$). While the age distribution for the incumbent peaks around 2 years, the age of the substitutes peaks earlier, corroborating the *recency* mechanism uncovered in Fig. 3E.

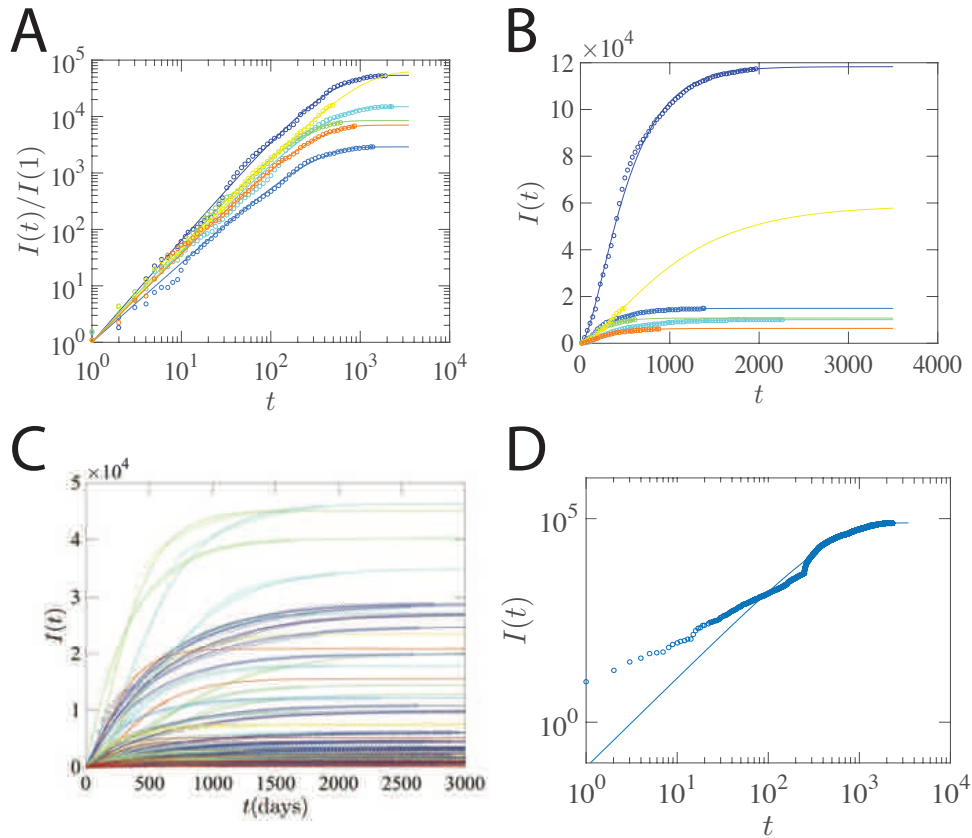

Supplementary Figure 23: **Model validation for handset dataset**. **(A–B)** Comparison between the empirical observation (open circle) and model simulation (solid line) for 6 randomly selected handsets. We show the normalized impact dynamics in (A) and the impact dynamics in (B). **(C)** Impact trajectories of 100 randomly selected handsets. **(D)** An example where the model fails to capture the growth patterns due to sudden shifts and jumps in impact dynamics.

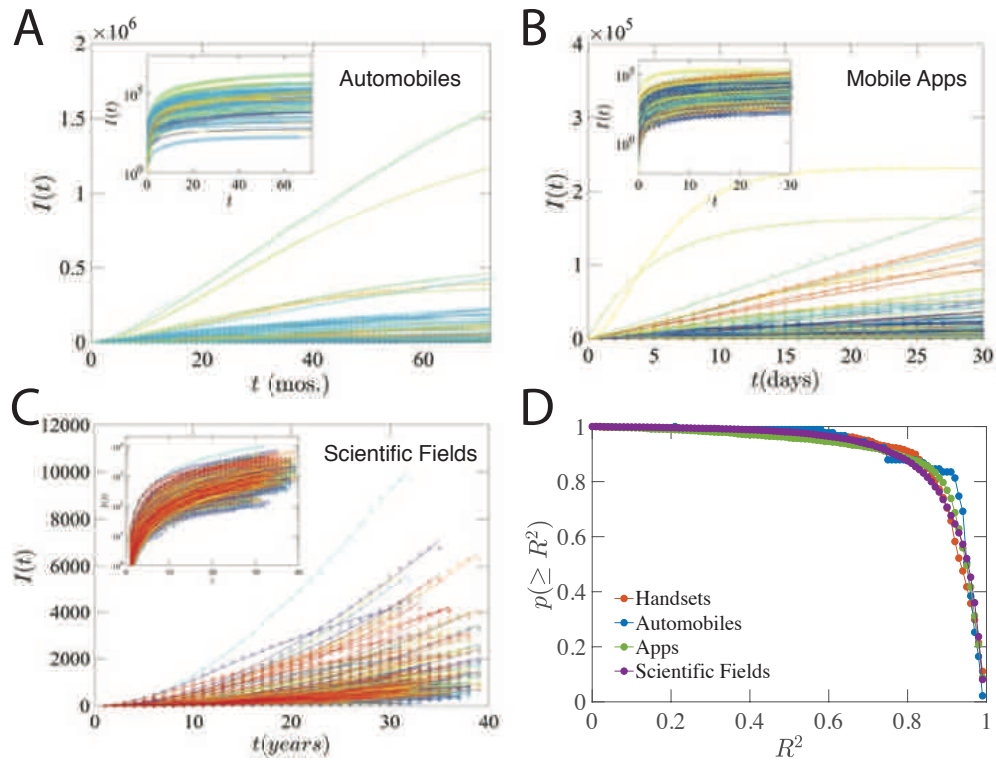

Supplementary Figure 24: **Model validation for automobiles, apps and scientific fields**. (**A–C**) Impact trajectories of 70 automobiles (A), 200 apps (B) and 500 scientific fields (C). (**D**) We fit each impact dynamics across four systems with the *MS* model and show the complimentary cumulative distribution of the $R^2$ of the fittings.

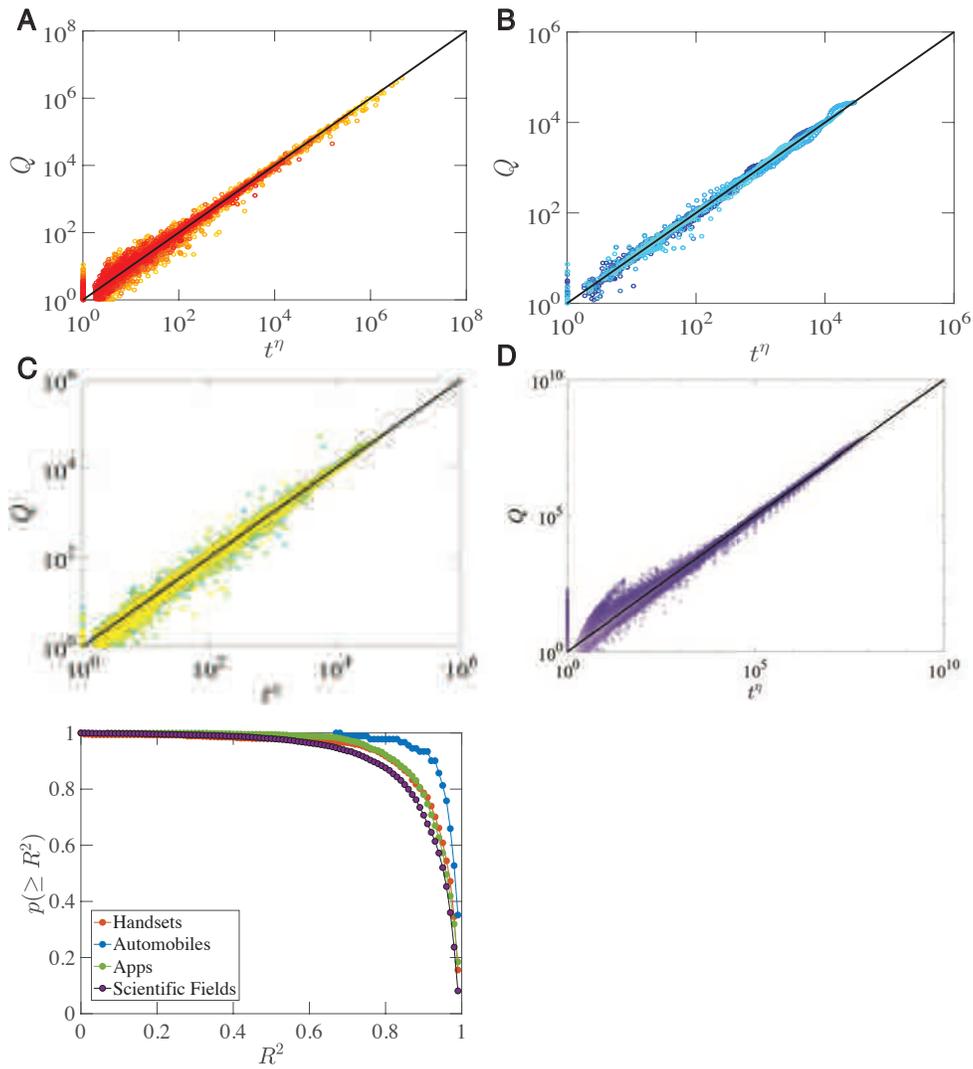

Supplementary Figure 25: **Rescaled Dynamics for Selected Products**. Rescaled impact $Q$ as a function of the rescaled time $t^\eta$ for selected (A) handsets, (B) automobiles, (C) mobile apps and (D) scientific fields with $R^2 > 0.9$. We also show the complementary cumulative distribution of the $R^2$ of the universal collapse (E), demonstrating that the rescaled impact dynamic for a vast majority of the products collapse into a universal curve.

70

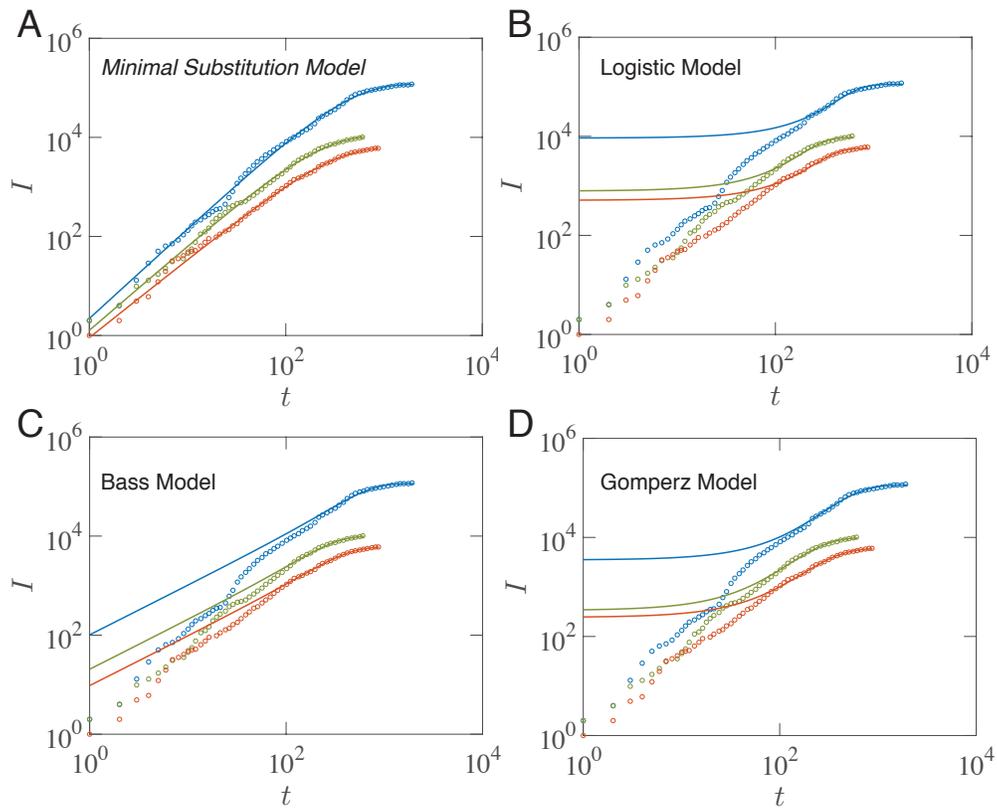

Supplementary Figure 26: **Handset Fitting Examples**. Fitting three handsets in the system as illustrative examples to highlight the conceptual differences between various models: **(A)** Minimal Substitution model, **(B)** Logistic model, **(C)** Bass model and **(D)** Gompertz model.

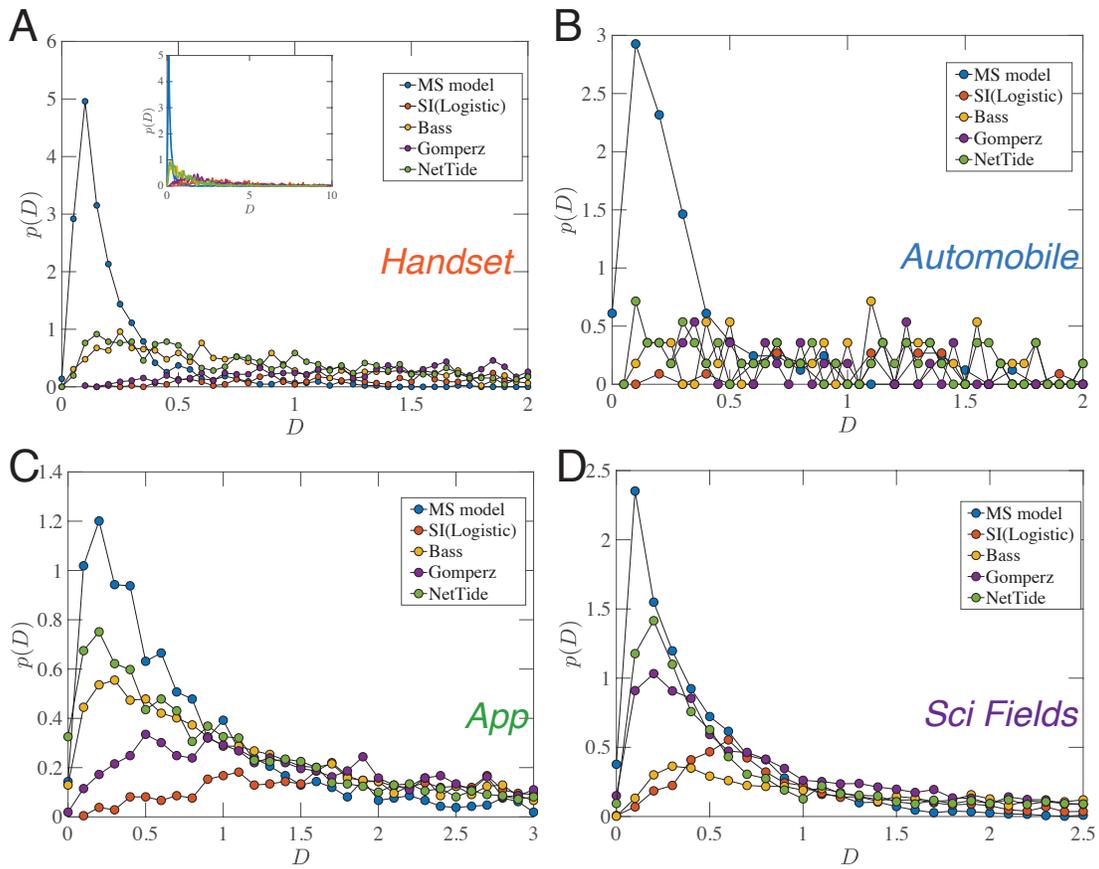

Supplementary Figure 27: Goodness of fit using weighted Kolmogorov-Smirnov (KS) test for (A) handsets, (B) automobiles, (C) smartphone apps and (D) scientific fields.

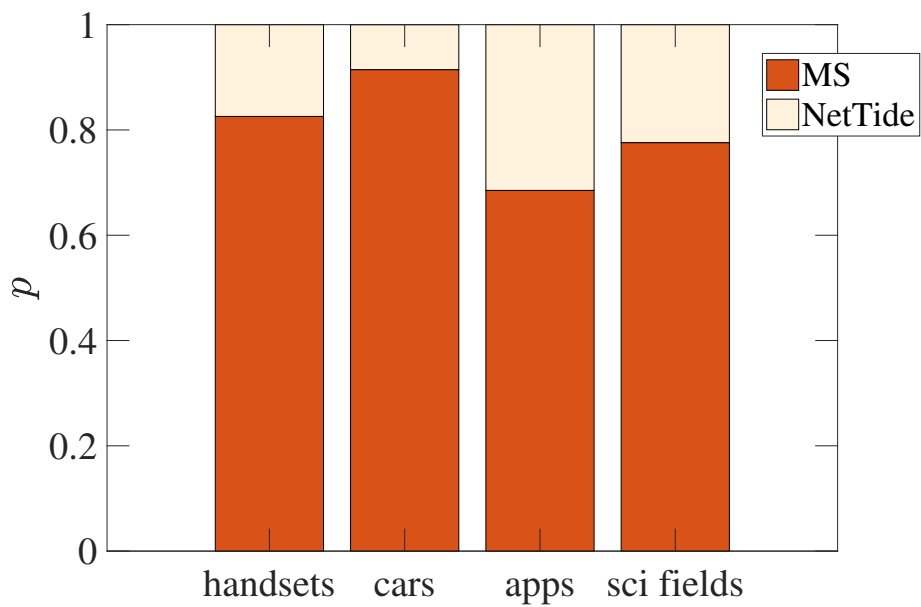

Supplementary Figure 28: Direct Comparison of the MS model and the NetTide model for four different datasets: handsets, automobiles, mobile apps and Scientific Fields. We adopt the weighted KS test to capture the goodness of fitting, where we find for 82.57% of handsets, 91.46% of automobiles, 68.52% of mobile apps and 77.59% of scientific fields, the MS model outperforms the NetTide model.

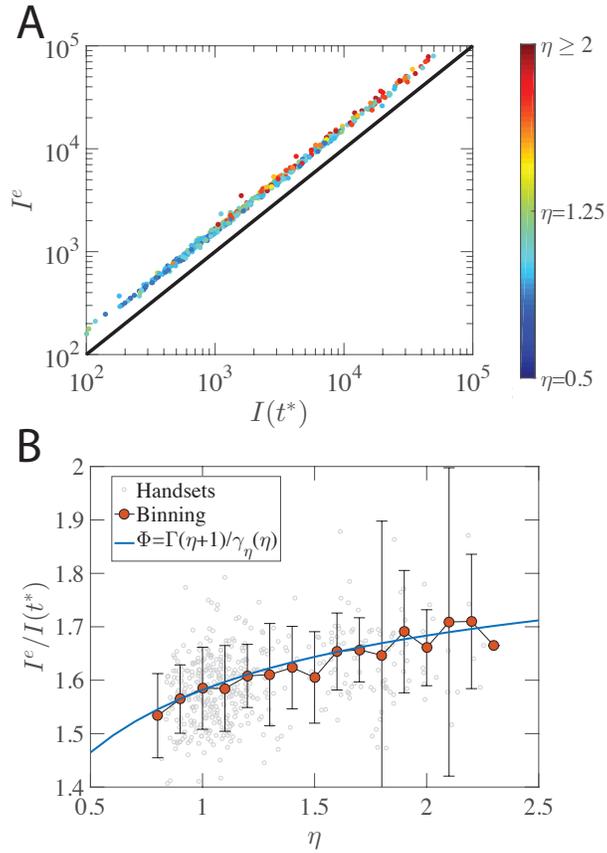

Supplementary Figure 29: **Relationship between short-term and long-term impact**. **(A)** $I^e$ as a function of $I(t^*)$. **(B)** The ratio between $I^e$ and $I(t^*)$ as a function of the fitness $\eta$. Solid line corresponds to the function $\Phi(\eta)$.

74

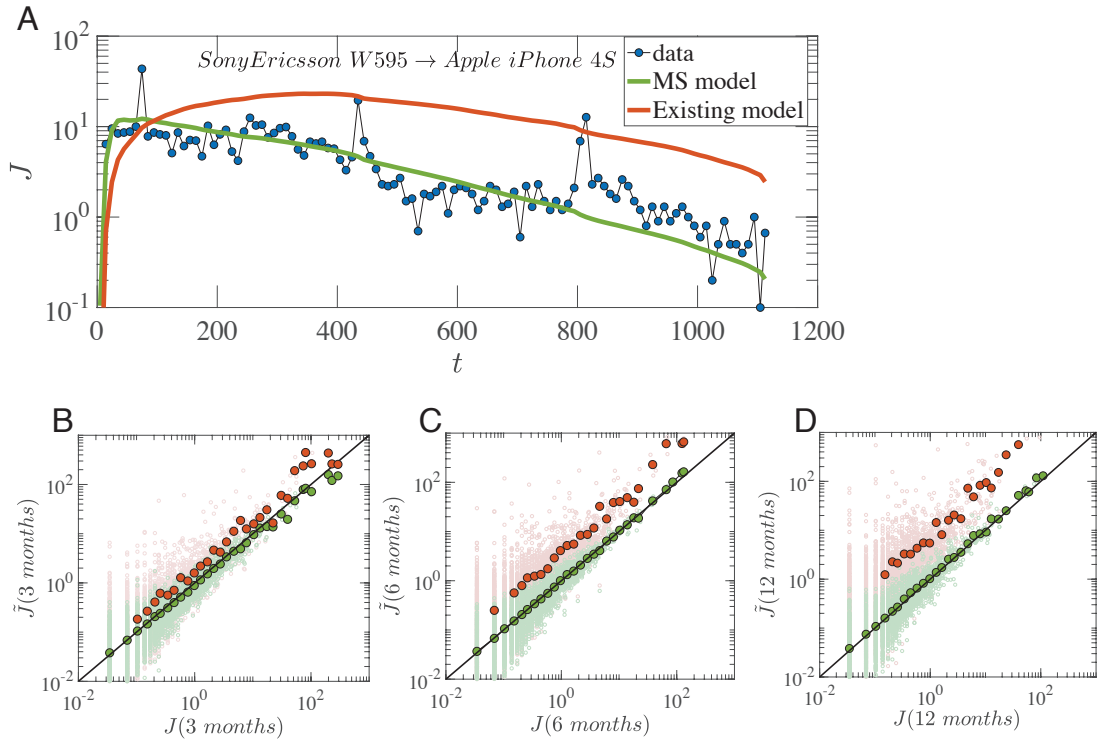

Supplementary Figure 30: **Quantifying substitution flows with *MS* model**. (A) The dynamic of the substitution flow from SonyEricsson W595 to Apple iPhone 4S. The green curve corresponds to the fit by MS model and the red curve by existing model. (B-D) By selecting three snapshots, we compare the model predicted substitution flow ($\tilde{J}$) with the amount in the real data ($J$), finding the MS model provides a rather good fit (Green Dots). The black line corresponds to $y = x$.

75

# Supplementary Tables

Supplementary Table 1: **Properties for various early growth patterns**

| Early Growth Pattern | Math form | Initial Impact | $n^{th}$ derivative for $t=0$ | Typical Models |
|---|---|---|---|---|
| Exponential | $I(t) = ae^{bt}$ | $a$ | $ab^n$ | Epidemic models (SIR or Logistic) |
| Linear | $I(t) = at$ | 0 | $a$ for $n=1$<br>0 for $n>1$ | Bass model |
| Power-Law with non-integer exponent | $I(t) = at^\eta$ | 0 | 0 for $n<\eta$<br>$\infty$ for $n>\eta$ | Minimal Substitution Model |

Supplementary Table 2: **Fitting early growth patterns to different functions**

| Function | Math form | Fitted parameters | Number of parameters |
|---|---|---|---|
| Exponential | $I(t) = ae^{bt}$ | $a, b$ | 2 |
| Linear | $I(t) = at$ | $a$ | 1 |
| Logistic | $I(t) = \frac{I^\infty}{1+e^{-k(t-t_0)}}$ | $k, t_0, I^\infty$ | 3 |
| Power-Law with non-integer exponent | $I(t) = at^\eta$ | $a, \eta$ | 2 |

Supplementary Table 3: **Early growth patterns of selected models** 1) For all models, $I(0)$ represents the initial impact of a given product. 2) For the SIR model, to avoid duplicate usage of letter, we use $A$ to represent number of current infected people. $S$ corresponds to the number of potential users and $R$ measures number of recovered people. The parameters satisfy the condition $S + A + R = N_0$. The impact of the product is captured by $I \equiv A + R$. 3) For the Flexible Logistic Growth model, $\mu$ and $k$ are constants and $t(\mu, k) = [(1+kt)^{\mu/k} - 1]/\mu$ for $\mu \neq 0$, $k \neq 0$, $t(\mu, k) = (1/k)log(1+kt)$ for $\mu = 0$, $k \neq 0$, $t(\mu, k) = (e^{\mu t} - 1)/\mu$, $\mu \neq 0$, $k = 0$, $t(\mu, k) = t$ for $\mu = 0$, $k = 0$.

| Model | Model Equation ($dI/dt =$) | Model Solution ($I =$) | Early Behavior ($I \sim$) |
|---|---|---|---|
| Logistic[5] | $qI(1 - I/I^\infty)$ | $I^\infty(1 + e^{-q(t-\tau)})^{-1}$ | $I(0)e^{qt}$ |
| Bass[41,42] | $(p + qI/I^\infty)(I^\infty - I)$ | $I^\infty \frac{1-e^{-(p+q)t}}{1+\frac{q}{p}e^{-(p+q)t}}$ | $I^\infty pt$ |
| Gompertz[43] | $qI \ln(I^\infty/I)$ | $I^\infty e^{-e^{-(a+qt)}}$ | $I(0)e^{e^{-a}qt}$ |
| SIR[30,31,49] | $\beta(I - R)(1 - I/I^\infty)$ | $N_0 - (N_0 - I_0)e^{-\frac{\beta}{\gamma}(R(t)/N_0)}$ | $I(0)e^{(\beta-\gamma)t}$ |
| Nelder[81] | $qI(1 - (I/I^\infty)^\phi)$ | $I^\infty(1 + e^{-\phi(c+qt)})^{-1/\phi}$ | $I(0)e^{qt}$ |
| Flexible logistic[82] | $q[(1+kt)^{1/k}]^{\mu-k}I(1 - I/I^\infty)$ | $I^\infty(1 + e^{-[c+qt(\mu,k)]})^{-1}$ | $I(0)e^{qt}$ |

# Supplementary References



\end{document}